\documentclass[11pt]{article}

\usepackage[preprint]{acl}

\usepackage{times}
\usepackage{latexsym}
\usepackage[T1]{fontenc}
\usepackage[utf8]{inputenc}
\usepackage{microtype}
\usepackage{booktabs}
\usepackage{array}
\usepackage{graphicx}
\usepackage{amsmath}
\usepackage{amssymb}
\usepackage{xurl}
\usepackage{enumitem}
\usepackage{xcolor} 
\usepackage{textcomp}

\newif\ifshownotes
\shownotestrue

\newcounter{note}[section]

\title{Behavioral Skill Reconstruction: \\ Cloning Hidden Functionality from LLM Agent Skills with \textsc{SkillClone}}

\author{
Peichun Hua \quad Haoxuan Xu \quad Mengyuan Li \\
\texttt{peichunhua04@gmail.com}\quad  \texttt{\{xuhaoxua,mli49061\}@usc.edu} \\
University of Southern California
}

\usepackage{xspace}
\newcommand{\method}{\textsc{SkillClone}\xspace}
\newcommand{\asr}{\mathrm{ASR}}

\begin{document}
\maketitle

\begin{abstract}

Closed-source agent skills encode proprietary instructions, scripts, constants, and data that providers expose as services while keeping the underlying packages hidden. Prior work focuses on prompt-injection attacks that directly disclose these artifacts, and existing defenses aim to block such leakage. File secrecy, however, does not stop users from recovering the underlying functionality through ordinary use.

We study behavioral skill reconstruction, in which an attacker submits valid task requests and uses the observed responses to build a functional clone of a hidden skill. We introduce \method, a black-box attack that forms an interface hypothesis from a skill's public advertisement, issues structured benign probes, synthesizes an executable clone, and iteratively repairs it through differential validation against the victim. Across 30 skills spanning rules, tables, procedures, and algorithms, \method recovers 19 of 21 mined skills above a trivial baseline on the strongest victims and reconstructs several procedures exactly, with disagreement-guided re-querying closing coverage gaps left by single-round synthesis. Because \method relies solely on legitimate interactions, disclosure filters, query-pattern detection, per-client query budgets, and advertisement minimization offer limited protection. Keeping skill files secret does not keep their functionality secret; defenses must also limit what accumulates through ordinary use. 
\end{abstract}

\section{Introduction}



LLM agents increasingly rely on external \emph{skills}, which package instructions, scripts, reference data, and tool logic that an agent loads on demand \citep{anthropic2025skills}. Skills add specialized capabilities without retraining the underlying model and improve task performance \citep{li2026skillsbench}. In a March 31, 2026 snapshot of skills.sh, \citet{wang2026skillstealing} counted 90{,}368 skills from 9{,}485 publishers and 24.3 million cumulative installs; they also cite Claw Mart, a paid marketplace with over 2{,}000 listings and \$100{,}000 in creator earnings. Providers therefore increasingly offer these capabilities as services while keeping the underlying packages confidential.

The commercial value of these skills creates an incentive to recover their capabilities without paying for or independently developing them. Existing research focuses on direct disclosure, where prompt injection, jailbreaking, or prompt stealing induces an agent to reveal hidden instructions or files \citep{zhang2024prompt,sha2024prompt,agarwal2024prompt,wang2026skillstealing}. Corresponding defenses detect extraction intent, isolate protected instructions, or filter outputs containing sensitive content \citep{hua2026rethinking,li2025drift,cao2025you,jiang2025promptkeeper}. This line of work treats skill confidentiality as artifact secrecy: a skill is considered protected as long as its files remain hidden.

Artifact secrecy, however, does not guarantee functional secrecy, because a skill deployed as a service reveals its underlying logic through its outputs. A navigation skill that ranks Polaris first among its preferred stars discloses that ranking whenever queried for its top choice, and a compensation skill applies its private $1.30\times$ benefits multiplier to every total it computes. Over repeated interactions, such responses expose table entries, thresholds, composition rules, and procedural conventions. We therefore ask: \textbf{\textit{Can an attacker reconstruct a skill's hidden functionality through ordinary task use while its artifacts remain concealed?}}

We formulate this threat as \textbf{behavioral skill reconstruction}, in which an attacker interacts with a skill-enabled agent solely through its service interface, submits valid task requests, and builds a functional clone without accessing or requesting the skill package. To study this threat, we introduce \method, a black-box attack that converts cues from the public skill \emph{advertisement} (used for task routing) into a typed interface hypothesis, issues structured benign probes, parses the responses, and synthesizes an executable clone. It then validates the clone on fresh inputs, identifies disagreements with the victim, and issues additional probes to repair the inferred implementation using only victim responses. To separate a skill's contribution from what the base model already knows, we define \emph{model-relative marginal functionality}: a component is \emph{IP-positive} and eligible for scoring only when the skill-enabled victim applies it reliably, whereas the same model cannot reproduce it from the advertisement alone.


We evaluate \method{} on 30 skills from SkillsBench~\citep{li2026skillsbench}, SkillRet~\citep{cho2026skillret}, and a public registry crawl, spanning rules, tables, procedures, and algorithms. In full-agent experiments, \method{} recovers 19 of the 21 mined skills above a trivial-baseline floor on the strongest victims and produces exact clones for several deterministic procedures. Controlled in-context experiments separate functional leakage from routing and framework failures and show that disagreement-guided re-querying closes gaps left by single-round synthesis. Existing defenses are largely ineffective: a disclosure filter that catches all four explicit disclosure probes, when enforced inside the attack loop, refuses only 4 of 5{,}508 queries and leaves ASR unchanged or higher on 18 of 21 skills. Query-pattern detection, per-client query budgets, and advertisement minimization similarly fail to prevent reconstruction.

Our contributions are:
\begin{enumerate}\itemsep0pt\parskip0pt\parsep0pt
\item We formulate behavioral skill reconstruction as a threat to closed-source agent skills and define model-relative marginal functionality to measure recovery beyond the base model.
\item We introduce \method, an iterative black-box attack combining public-advertisement hypotheses, structured benign probes, executable synthesis, and differential validation.
\item We evaluate \method{} on 30 skills, identify six reconstruction bottlenecks, and test four defense interventions.
\end{enumerate}

\section{Background}
\label{sec:threat}

\paragraph{Agent Skills.} A \emph{skill} is a reusable, file-system-based package that an agent loads during task execution to handle a class of tasks without retraining the model \citep{anthropic2025skills,li2026skillsbench}, building on prior work on reusable agent capabilities \citep{wang2023voyager,wang2024agentworkflowmemory,xu2026agentskillssurvey,liu2026skillsvote}.

We model a skill as $s=(a_s,b_s)$. The \emph{advertisement} $a_s$ is the public name and description used for routing \citep{cho2026skillret}; the \emph{body} $b_s$ contains the full instructions and any bundled scripts, data, or resources. Under progressive disclosure \citep{anthropic2025skills}, an agent initially sees $a_s$, loads $b_s$ when selected, and opens auxiliary files as needed. Mounting $s$ on model $m$ yields an agent $A_{m,s}$ whose observable behavior depends on the hidden body. \method{} exploits the fact that $a_s$ exposes the routing interface while ordinary responses reveal the body's functional behavior.

\paragraph{Related Studies.} Prior work on prompt confidentiality targets the hidden text that conditions a model: adversarial queries can reconstruct system prompts and in-context instructions \citep{zhang2024prompt,sha2024prompt,agarwal2024prompt}, and skill-stealing attacks can induce an agent to print its \texttt{SKILL.md} \citep{wang2026skillstealing}. More recent studies examine agent-side leakage through tool interfaces, including exfiltration from memory retrieval \citep{liu2026spore}, black-box distillation of multi-agent orchestration harnesses \citep{cui2026harness}, and skill-ecosystem risks addressed by malicious-skill detection, runtime sandboxing, and composition auditing \citep{kim2026skillmutator,lan2026runtime,xie2026composition,hossain2026skillvetbench}. We instead measure how much of a benign skill's functionality leaks through task-valid use.

Concurrently, two works reconstruct skill behavior without disclosure. \textsc{SigLeak} \citep{geng2026sigleak} contrasts skill-enabled and skill-disabled execution trajectories to infer a skill's contribution. \textsc{Daydreaming} \citep{tsai2026daydreaming} keeps the skill enabled and observes only final answers using adaptive shadow-agent probing to recover 86.8\% of skill capability at a median of 32 victim queries across 7 skills. \method{} evaluates a larger and more varied suite of 30 skills from three sources (21 mined from public registries, 9 from SkillsBench) spanning data, rules, and procedures across five victims from four model families, scores each clone by exact agreement with a deterministic oracle on held-out inputs, and traces failures to six bottlenecks (victim fidelity, probe specificity, coverage, hypothesis quality, synthesis, and recall) via targeted interventions. \textsc{CapaTheft} \citep{lyu2026capatheft} complements our setting: it transfers a victim's capability (typically from a stronger model) to a weaker agent given a known task instance and the victim's successful trajectory, whereas \method{} recovers the skill from final answers alone.


\section{Attack Methodology}
\label{sec:method}

\begin{figure*}[t]
\centering
\includegraphics[width=0.96\textwidth]{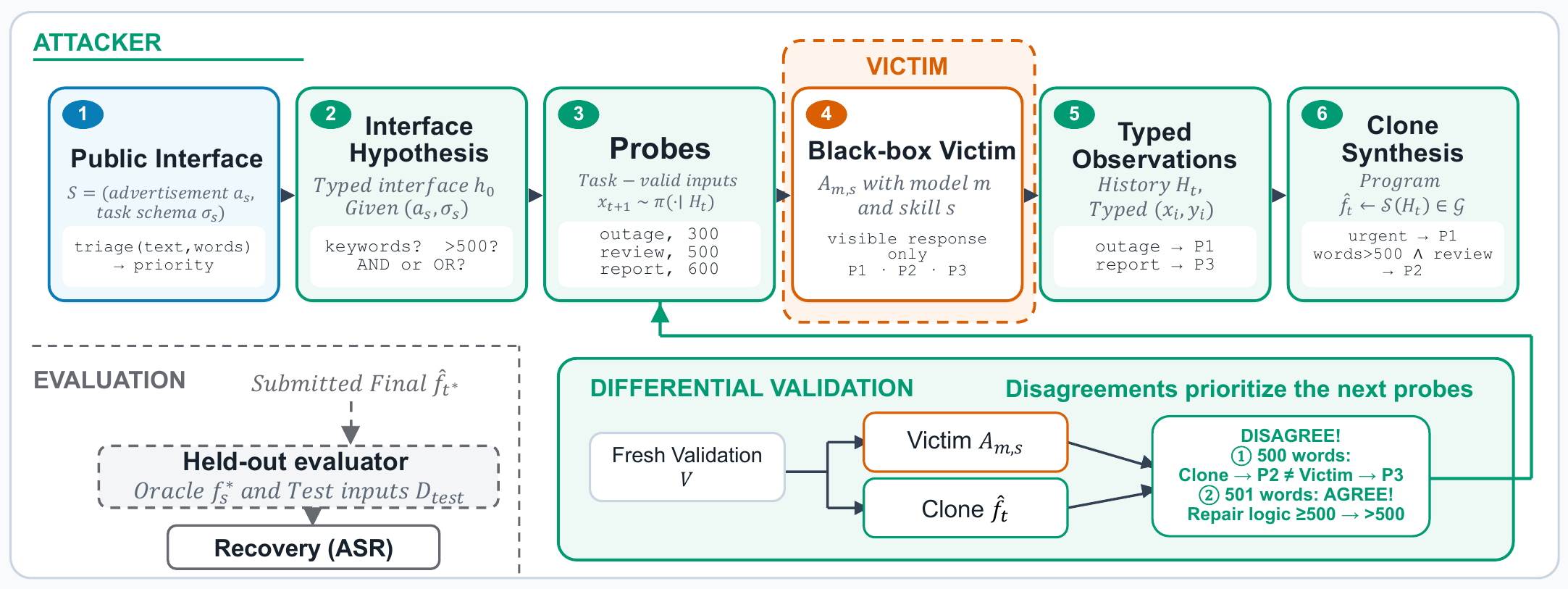}
\caption{Overview of \method{}. Stages 1--3 turn the public advertisement and task interface into an interface hypothesis and diverse probes. Stages 4--6 send probe inputs to the victim, parse its responses, and synthesize the clone. Differential validation compares clone and victim outputs and prioritizes disagreements for the next round.}
\label{fig:skillclone-overview}
\end{figure*}

We formulate \method{} as black-box system identification of the functionality contributed by a mounted skill (Figure~\ref{fig:skillclone-overview}), defining the threat model (\S\ref{sec:formal-obj}), target model (\S\ref{sec:component-model}), and five attack modules (\S\ref{sec:modules}).

\underline{\textit{Running example.}}
To illustrate \method, we use a synthetic document-triage skill that assigns text to three priority levels (1--urgent, 2--review, and 3--normal). Its advertisement specifies content and word count as inputs, while its body encodes an urgent-keyword disjunction, a length threshold ($>500$ words), and a conjunction requiring both high word count and review-related content for Priority~2.

\subsection{Threat Model and Formal Objective}
\label{sec:formal-obj}
Following prior work \citep{wang2026skillstealing,cui2026harness}, we adopt a black-box threat model in which the attacker interacts with a skill-enabled agent as an ordinary user. Given only the advertisement $a_s$ and public task interface $\sigma_s$, the attacker submits \emph{task-valid} inputs and observes only final messages and output artifacts; the skill body, bundled files, hidden tool activity, chain-of-thought traces, and evaluator test cases remain inaccessible. The attacker's goal is an executable clone $\hat{f}$ that reproduces the functionality contributed by the mounted skill.

Each query must request an output that the advertised skill is intended to provide; artifact-revelation and instruction-exfiltration queries fall outside the protocol. This restriction separates \method{} from jailbreak, prompt-injection, and prompt-extraction settings \citep{wang2024adashield,hua2026rethinking,li2025drift,cao2025you,jiang2025promptkeeper}. Although all evaluated skill bodies are public artifacts available to the evaluator, the protocol withholds them from the attacker.

Within this threat model, we formalize the reconstruction objective as follows.
For a victim model $m$ and skill $s$, let $A_{m,s}$ denote the agent exposed to users. A task input $x\in\mathcal{X}$ produces an observable answer $y=A_{m,s}(x)$ consisting only of visible messages and artifacts. The attacker receives the advertisement $a_s$, public task interface $\sigma_s$, and query budget $B$, and its information after $t$ queries is
\begin{equation}
\begin{aligned}
H_t &= \{a_s,\sigma_s, (x_1,y_1), \ldots, (x_t,y_t)\},\\
y_i &= A_{m,s}(x_i).
\end{aligned}
\end{equation}
The attacker's policy chooses $x_{t+1}\sim\pi(\cdot\mid H_t)$ subject to the task-valid restriction above, and a synthesizer maps the resulting history to an executable clone in a class $\mathcal{G}$. The evaluator alone holds the skill's ground-truth function $f_s^\star$ and uses it to score the final clone (\S\ref{sec:evaluation}); the attacker optimizes agreement with victim responses.


\subsection{Component-oriented Target Model}
\label{sec:component-model}


The reconstruction target is the \emph{model-relative marginal functionality} contributed by the hidden skill body: concrete thresholds, table entries, composition rules, and procedural conventions that distinguish the skill-enabled agent from the base LLM. The public advertisement and task interface cue the likely implementation structure without revealing its exact parameters: a pricing skill suggests a lookup table indexed by products or service tiers, whereas a traffic classifier suggests threshold rules over network features. From these cues, the attacker forms an initial typed interface hypothesis $h_0$, and probing and synthesis recover the concrete constants, entries, and logic.

\subsection{Autonomous Attack Modules}
\label{sec:modules}
\method{} consists of five modules connected by their data flow. Interface hypothesis formation narrows the space of plausible implementations; probe generation and observation parsing convert task-valid interactions into typed evidence; program synthesis produces an executable clone; and differential validation uses victim--clone disagreements to guide repair.

\paragraph{Interface hypothesis.} Given $(a_s,\sigma_s)$, the attacker constructs a typed interface hypothesis
\begin{equation}
h_0=(\hat{\mathcal{X}},\hat{\mathcal{Y}},\hat{Z},\hat{\Theta},\hat{\mathcal{C}}),
\end{equation}
where $\hat{\mathcal{X}}$ and $\hat{\mathcal{Y}}$ are candidate input and output types, $\hat{Z}$ are candidate latent factors, $\hat{\Theta}$ are candidate constants or parameters, and $\hat{\mathcal{C}}$ is a candidate composition family. Because $h_0$ relies on public cues and general domain knowledge, unadvertised factors outside that knowledge may be omitted, creating a hypothesis-quality bottleneck. In the document-priority example, public information exposes the input fields, output labels, and the relevance of keywords and length, while leaving the exact keywords, threshold, and composition unknown.

\paragraph{Probe policy.} The attacker uses a fixed, domain-general operator library to discriminate among competing hypotheses: \emph{isolation} varies one factor while holding others neutral; \emph{dose--response} sweeps its magnitude; \emph{boundary search} localizes a threshold; \emph{composition} distinguishes additive, max/min, priority, floor, and Boolean rules; \emph{counterfactual} and \emph{confound} probes separate overlapping factors; and \emph{enumeration} covers lookup keys. The attacker instantiates these heuristic experimental-design operators from the public interface and observed history.


\textit{\underline{Running example.}} The attacker applies three operators:

\textit{Isolation.} To test whether specific keywords trigger urgent priority, the attacker holds length constant at a neutral 300-word body and submits ``server outage'' and ``server reboot.'' Both yield Priority~1, indicating an urgent-keyword trigger list.

\textit{Boundary search.} To localize a suspected word-count threshold for the review tier, the attacker submits ``budget review'' at 500 and 501 words, receiving Priority~3 and Priority~2, respectively, which identifies a strict $>500$-word threshold.

\textit{Composition.} To test whether length alone suffices, the attacker submits a neutral 600-word ``quarterly report,'' which still returns Priority~3, confirming that Priority~2 requires both long text \textit{and} review-related content. 

\paragraph{Observation parsing.} Because real agents return prose and artifacts, the attacker requests ``only the answer,'' parses each visible response into a typed observation, and treats the parsed output as the training label for synthesis and repair.

\paragraph{Program synthesis.} The synthesizer $\hat{f}_t=\mathcal{S}(H_t)\in\mathcal{G}$ infers private constants and logic from the query history to produce a standalone executable clone matching a constrained schema appropriate to the hypothesized interface (threshold classifier, table lookup, rule engine, state machine, or numeric procedure).

\underline{\textit{Running example.}} From observations such as (``server outage'', 300 words) $\to$ Priority~1, (``budget review'', 500 words) $\to$ Priority~3, (``budget review'', 501 words) $\to$ Priority~2, and (``quarterly report'', 600 words) $\to$ Priority~3, the synthesizer infers three rules: (1) urgent keywords trigger Priority~1 regardless of length; (2) Priority~2 requires $>500$ words and review-related content; and (3) all other inputs default to Priority~3. It then emits a Python function encoding these branches.

\paragraph{Differential validation and repair.} Because the reference oracle remains hidden, the attacker evaluates candidates by agreement with the victim. It generates fresh task-valid inputs, executes both the synthesized clone $\hat{f}_t$ and $A_{m,s}$, re-queries disagreements, and synthesizes a repaired clone.

\underline{\textit{Running example.}} If the initial clone uses $\mathrm{words}\geq500$, ``budget approval'' at exactly 500 words produces a disagreement: the clone returns Priority~2 whereas the victim returns Priority~3, while both return Priority~2 at 501 words. These observations isolate the strict ${>}500$ comparison and guide the repair. The attacker returns the candidate $\hat{f}_{t^*}$ achieving the highest victim agreement on its validation set:
\begin{equation}
t^*=\arg\max_t \; \mathrm{Agree}(\hat{f}_t,A_{m,s};V),
\end{equation}
where $V$ is generated by the attacker and is disjoint from the evaluator test data.

\subsection{Reconstruction Loop}

Putting the modules together, one round updates the interface hypothesis from $H_t$, instantiates a discriminating probe, parses the victim response, synthesizes $\hat{f}_{t+1}$, and validates it against fresh victim outputs. The loop repeats until budget $B$ is exhausted and returns the round with the highest victim agreement. This modular design enables ablations that replace one module at a time and measure the resulting change in ASR (\S\ref{sec:results-bottlenecks}).

\section{Experimental Setup}
\label{sec:evaluation}
We first construct the skill suite and determine eligible skill--victim cells (\S\ref{sec:suite}), then build component-level held-out evaluators (\S\ref{sec:metrics}) and specify the \method{} protocol (\S\ref{sec:protocol}).

\subsection{Eligibility and Suite Construction}
\label{sec:suite}

Our suite enables exact measurement of behavioral reconstruction by requiring deterministic logic, auditable ground truth, and measurable skill benefit on at least one victim.

\paragraph{Candidate pools and execution settings.}
We draw candidates from SkillsBench \citep{li2026skillsbench}, SkillRet \citep{cho2026skillret}, and a public registry crawl. Within each cohort, eligibility screening and \method{} evaluation use the same execution setting. SkillsBench candidates are evaluated in the benchmark's native OpenHands environment because these packages are curated for agent execution, whereas mined candidates are screened and evaluated by loading the skill body directly in context to isolate \method{} from agent-stack delivery and parsing failures. Appendix~\ref{app:deployed} additionally reports full-agent eligibility screens and complete \method{} runs for the mined skills.

\paragraph{Victim models.}
We evaluate DeepSeek-V4-Flash (Flash), DeepSeek-V4-Pro (Pro), GLM-5.1 (GLM) \citep{glm2026}, Kimi-K2.6 (Kimi) \citep{moonshot2026kimik26}, and GPT-5.6-Luna (Luna) as victims, defining the skill--victim cells used in the second screening stage and cross-model evaluation.

\paragraph{Two-stage screening.}
The first stage retains targets with deterministic logic and auditable ground truth; mined candidates additionally require a compatible license. In the second stage, we evaluate each surviving candidate with and without the skill body on identical task-valid inputs for each victim. A component is \emph{marginal} when $A_{m,s}$ applies it reliably and $A_{m,\varnothing}$ fails to do so; a skill--victim cell is IP-positive if it contains at least one marginal component. Operationally, we require $\mathrm{fid}(A_{m,s})\ge 0.6$ and $\mathrm{fid}(A_{m,s})-\mathrm{fid}(A_{m,\varnothing})\ge 0.25$, where $\mathrm{fid}$ denotes agreement with the oracle.

A skill enters the suite if at least one victim yields an IP-positive cell, and reconstruction for a given victim is evaluated only on its eligible cells. This procedure yields a 30-skill suite comprising 9 code-execution or procedure-heavy SkillsBench skills, 11 SkillRet data/rule skills, and 10 data/rule skills from the registry crawl (sources, license filter, and per-victim counts in Appendix~\ref{app:benchmark}).

\subsection{Held-out Evaluator and Scoring}
\label{sec:metrics}

For each retained target, we build an evaluator-only harness comprising a deterministic reference oracle $f_s^\star$ reimplemented from the skill body, a typed input/output schema, and a seeded input generator. The evaluator decomposes each skill into functional components $c\in C_s$ and constructs discriminating inputs $D(c)$ on which correct and incorrect implementations diverge, exercising hidden constants, thresholds, table rows, special cases, composition rules, and procedural conventions.

Test inputs are disjoint from queried inputs wherever the domain permits; on small finite domains (e.g., \texttt{delegation}'s 7 inputs) they cover the full domain, so ASR measures recovery of the entire function. Mined-skill clones are scored on $n{=}150$ evaluator-only test inputs, and SkillsBench clones on $n{=}800$. The test distribution emphasizes discriminating inputs so that held-out agreement reflects recovery of hidden components rather than agreement on uninformative examples. Appendix~\ref{app:benchmark} details the per-skill oracles, component definitions, and split construction.

The primary metric is attack success rate:
\begin{equation}
\asr(\hat{f})=
\mathbb{E}_{x\sim\mathcal{D}_{\mathrm{test}}}
\left[\mathbf{1}\{\hat{f}(x)=f_s^\star(x)\}\right].
\end{equation}
The deployed clone is selected by victim agreement on attacker-generated validation inputs, whereas reported ASR is computed on held-out inputs against $f_s^\star$.

Each skill's \emph{floor} is the ASR achieved by a trivial baseline clone on the same held-out test set: a majority-class predictor for categorical outputs, or an all-zero or identity clone when no majority class applies (e.g., \texttt{r2r}). A skill is considered \emph{above floor} only when its ASR exceeds this baseline. Table~\ref{tab:mined} (Appendix~\ref{app:extra}) reports each skill's floor, and Appendix~\ref{app:benchmark} details two rebalancing exceptions.

\subsection{\method{} Experimental Protocol}
\label{sec:protocol}

\paragraph{Attacker model.}
DeepSeek-V4-Flash (Preview) \citep{deepseekai2026deepseekv4} is the primary attacker throughout our experiments; we use Pro only in the attacker-model ablation (Appendix~\ref{app:attacker-model}).

\paragraph{Query protocol.} Our main results use a paired protocol of 162 victim queries per skill: 12 initial probes, four rounds of 30 candidate inputs, and 30 validation inputs. In each round, the \emph{adaptive} policy adds the 10 answered candidates with the largest clone--victim disagreements to the training set (at most 52 training answers in total), while the remaining 80 candidates are used only for ranking and the 30 validation answers select the best round. As a controlled ablation, a \emph{fixed} policy draws 10 random candidates per round from the same answered pool; we report the adaptive policy throughout. The cross-victim comparison (Figure~\ref{fig:cross-victim-results}, Appendix~\ref{app:extra}) uses an 82-query sequential variant that queries only its training inputs (12 initial probes, four rounds of 10 inputs, and 30 validation inputs). SkillsBench attacks use $K{=}24$ probes and 30 validation queries (54 total). Synthesis with Flash costs roughly \$0.20 per skill for the adaptive attack.

\paragraph{Model and synthesis settings.}
Flash and Pro, as victims and as attacker, run in DeepSeek's default thinking mode, which samples with fixed settings and ignores the temperature argument \citep{deepseek2026thinking}. GLM-5.1 runs at $T{=}0$ with thinking disabled, Kimi-K2.6 at $T{=}1$ in its default thinking mode \citep{moonshot2026kimik26}, and Luna exposes no temperature parameter. The synthesizer draws three independent candidates and retains the one with the highest victim-validation agreement. Each synthesis call has a 32k-token budget (extended to 64k if the response is truncated), and a candidate that fails to compile is allowed up to three repair attempts. A round in which no candidate compiles falls back to the majority-answer clone and is treated as an attack failure.
Appendices~\ref{app:deployed} and~\ref{app:temp} report full-agent parsing and robustness to victim sampling.

\section{Results}
\label{sec:results}


We evaluate whether \method{} reconstructs hidden functionality and what limits recovery (\S\ref{sec:results-bottlenecks}), what the victim queries contribute beyond the attacker model's knowledge (\S\ref{sec:queries}), how closed-loop refinement and adaptive probing help (\S\ref{sec:closed-loop}), and how results vary across victims (\S\ref{sec:cross-model}).

\subsection{Reconstruction Effectiveness}
\label{sec:results-bottlenecks}

\paragraph{Reconstruction by component type.} Table~\ref{tab:taxonomy} groups the 30 skills by the kind of functionality they hide and gives one real hidden rule of each kind; source-decomposed and per-skill results are in Appendices~\ref{app:same-setting} and~\ref{app:extra}.

\begin{table*}[t]
\centering
\small
\setlength{\tabcolsep}{4pt}
{%
\begin{tabular}{lc>{\raggedright\arraybackslash}p{6.6cm}ccl}
\toprule
Component type & $n$ & A hidden rule of this type & Median ASR & $>$floor & Bottleneck \\
\midrule
Threshold rules & 5 & port scan if entropy $>6.0$ bits, SYN ratio $>0.7$, $>100$ ports (\texttt{dapt}) & 84\% & 4/5 & probe specificity \\
Lookup tables & 10 & Kling AI credit price for each video and image tier (\texttt{klingai-pricing}) & 85\% & 8/10 & coverage \\
Rule composition & 3 & an adjacent Government Plaza adds $+1$ to specialty districts (\texttt{civ6}) & 87\% & 3/3 & hypothesis quality \\
Procedures & 6 & work before 4\,AM belongs to the previous day (\texttt{working-day}) & 100\% & 5/6 & synthesis \\
Numeric formulas & 6 & total pay $=1.30\times$ base salary $+$ equity (\texttt{team-composition}) & 45\% & 3/6 & recall \\
\bottomrule
\end{tabular}%
}
\caption{Full-agent extraction ASR by dominant component type. Each skill contributes its median ASR across available full-agent (OpenHands) victims: Flash, Pro, and Luna for mined skills, and all five victims for SkillsBench. Pooled median and $>$floor fraction are computed across skills per type.}
\label{tab:taxonomy}
\end{table*}

\paragraph{Bottleneck diagnostics.} We isolate each failure mode via targeted interventions (Table~\ref{tab:mechanisms}, Appendix~\ref{app:spectrum}): replacing victim labels with reference-oracle labels tests \emph{victim fidelity}; hand-designed boundary sweeps test \emph{probe specificity}; larger query budgets test \emph{coverage}; supplying an unadvertised latent factor tests \emph{hypothesis quality}; providing a named algorithm or step-by-step reasoning (Appendix~\ref{app:recipe-extraction}) tests \emph{synthesis}; and supplying a formula's name tests \emph{recall}.

\textbf{Victim fidelity} bounds every component type because victim responses are the attacker's only labels. On \texttt{r2r}, the victim miscomputes the Jacobian entries encoding hidden constants; the clone faithfully reproduces those errors and scores 0\%, whereas reference-oracle labels raise ASR to 100\%.

\textbf{Threshold rules} fail when no probe lands near the decision boundary. Coarse autonomous probes separate the categorical constraints of \texttt{security-environment-standards} but miss \texttt{dapt}'s entropy and packet-rate cutoffs, leaving \texttt{dapt} at its 50\% floor; hand-designed boundary sweeps raise it to 96--100\%.

\textbf{Lookup tables} are limited by coverage. The attacker enumerates the few dozen tiers of \texttt{klingai-pricing}, but 24 probes cover only 3 of \texttt{codebook}'s 230 entries (0\% ASR); coverage-limited skills improve with larger query budgets (Appendix~\ref{app:probe-ablation}).

\textbf{Rule composition} fails when an unadvertised interaction is omitted from the hypothesis. On \texttt{civ6}, ASR plateaus near 60\% because the attacker never hypothesizes that an adjacent Government Plaza adds $+1$; once this factor is supplied, recovery reaches 100\%.

\textbf{Procedures} are the most tractable category for synthesis. Clones of \texttt{drone}, \texttt{working-day}, and \texttt{reflow} match every held-out case, and the S3-FIFO state machine behind \texttt{cache} reaches 99.8\%. The exception is \texttt{osm-topology}'s modular hue arithmetic; prompting the victim to output intermediate reasoning steps (``recipe extraction,'' Appendix~\ref{app:recipe-extraction}) lifts recovery to 100\%.

\textbf{Numeric formulas} are primarily limited by formula recall. Although \texttt{labor-rate} and \texttt{protein-qc-score} reach 100\%, \texttt{powerlifting} remains near 0\% despite faithful victim responses because the synthesizer repeatedly fits the better-known Wilks formula rather than DOTS; supplying the formula name resolves the gap completely.

\paragraph{Clone structure.}
Successful clones achieve 74--100\% ASR while sharing only 35--59\% of their tokens with the skill body, confirming that they reimplement the functionality rather than reproducing text (Appendix~\ref{app:clone-structure}).

\subsection{Contribution of Victim Queries}
\label{sec:queries}

We quantify what victim responses add beyond the attacker model's prior knowledge using three controls on the 21 mined skills (Pro victim, controlled setting; Table~\ref{tab:queries}, per-skill results in Appendix~\ref{app:zeroquery}).

\begin{table}[t]
\centering
\small
\resizebox{\columnwidth}{!}{%
\begin{tabular}{lcccc}
\toprule
Component ($n$) & Floor & Zero-query & Plain LLM & \method{} \\
\midrule
Rule composition (2) & 19 & 0 & 39 & 100 \\
Numeric formulas (4) & 4 & 2 & 52 & 94 \\
Lookup tables (8) & 22 & 19 & 18 & 68 \\
Threshold rules (4) & 48 & 43 & 37 & 59 \\
Procedures (3) & 14 & 68 & 77 & 77 \\
\midrule
All (21) & 22 & 25 & 39 & 75 \\
\bottomrule
\end{tabular}%
}
\caption{Mean ASR (\%) by component type (21 mined skills, Pro victim). Zero-query: clone written from the advertisement alone (mean of five runs). Plain LLM: attacker LLM selects its own queries within 162 queries.}
\label{tab:queries}
\end{table}

\begin{figure*}[t]
\centering
\footnotesize
\setlength{\tabcolsep}{5pt}
\begin{tabular}{>{\raggedright\arraybackslash}p{0.20\textwidth}>{\raggedright\arraybackslash}p{0.33\textwidth}>{\raggedright\arraybackslash}p{0.37\textwidth}}
\toprule
\textbf{Hidden in the skill body} & \textbf{Attacker's guess with no queries} & \textbf{\method{} clone after querying the victim} \\
\midrule
\multicolumn{3}{l}{\texttt{stars-we-prefer}: ``our 10 preferred stars for celestial navigation''} \\[2pt]
1.\ Polaris, 2.\ Sirius, 3.\ Betelgeuse, 4.\ Vega, 5.\ Rigel, 6.\ Arcturus, 7.\ Capella, 8.\ Aldebaran, 9.\ Antares, 10.\ Deneb
&
\texttt{\# the ten brightest stars}\newline \texttt{1:"Sirius", 2:"Canopus",}\newline \texttt{3:"Rigil Kentaurus", 4:"Arcturus", ...}\newline ASR 0\%
&
\texttt{\# rank-to-star mapping from probes}\newline \texttt{1:"Polaris", 2:"Sirius", 3:"Betelgeuse",}\newline \texttt{4:"Vega", ..., 10:"Deneb"}\newline ASR 100\% \\
\midrule
\multicolumn{3}{l}{\texttt{team-composition}: ``compute total compensation from base salary and equity value''} \\[2pt]
Total Comp $=$ Base Salary $\times$ 1.30 (benefits \& taxes) $+$ Equity Value
&
\texttt{return base\_salary + equity\_value}\newline ASR 0\%
&
\texttt{return 1.3 * base\_salary + equity\_value}\newline ASR 100\% \\
\bottomrule
\end{tabular}
\caption{What victim queries contribute. From the advertisement alone, the attacker LLM falls back on public knowledge: the brightest stars and a plain salary sum. After querying the victim, \method{} recovers the skill's private choices---Polaris ranked first and a $1.30\times$ benefits multiplier (Pro victim, controlled setting; held-out ASR).}
\label{fig:clone-examples}
\end{figure*}

\paragraph{Beyond the attacker LLM.} A plain-LLM baseline receives the same advertisement, schema, and 162-query budget, proposes its own inputs, and synthesizes a clone from up to 132 answers. Using only 52 training answers, \method{} achieves 75.3\% mean ASR versus 38.5\%, outperforms the baseline by more than 5~pp on 14 of 21 skills, and trails on none. Although the plain LLM collects plausible answers, it rarely translates them into a working clone; on \texttt{labor-rate}, it scores 1\% compared to \method{}'s 88\%.

\paragraph{Public versus proprietary functionality.} Skills grounded in public knowledge can often be reconstructed from their advertisements alone: \texttt{token-cost-tracking} multiplies token counts by published API prices and averages 82\% without queries. Skills encoding provider-specific choices cannot (Figure~\ref{fig:clone-examples}): \texttt{labor-rate} combines custom trade wages, regional multipliers, and burden markups, and \texttt{team-composition} applies a $1.30\times$ benefits load; both score 0\% in every zero-query run, whereas \method{} recovers them to 88\% and 100\%. The 16 skills below 50\% zero-query ASR rise from 9.6\% to 68.3\% on average, and the remaining 5 from 75.9\% to 97.5\%.

\paragraph{Altered skills.} To rule out memorization, we modify hidden constants and rules in both the skill body and the oracle so that no public source matches the victim (e.g., shifting \texttt{working-day}'s day boundary from 4~AM to 10~AM). \method{} adapts to the altered behavior and beats the zero-query baseline on 10 of 11 altered skills, lifting \texttt{team-composition} from 0\% to 100\% and \texttt{working-day} from 62\% to 99\% (Appendix~\ref{app:zeroquery}).

\paragraph{Rule-level recovery.} To verify that held-out ASR reflects genuine rule recovery, we measure statement and path coverage of the test suite against each reference oracle. Test inputs execute 73--100\% of reference statements, with uncovered lines limited to input-validation errors and single-input special cases. Averaging accuracy across execution paths weights rare rule combinations equally: clones are exact on all 6 paths of \texttt{bellog-structure} and on \texttt{working-day}'s before-4-AM exception, and across \texttt{rbac-validator}'s 24 role--permission paths, per-path accuracy (90.6\%) closely tracks overall ASR (92.7\%) (Appendix~\ref{app:zeroquery}).

\subsection{Effects of Closed-loop Refinement}
\label{sec:closed-loop}

Closed-loop refinement adds 25--30~pp on every victim across the 13 skills whose mean lift exceeds 5~pp (Appendix~\ref{app:close-loop-refine}), bringing \texttt{stars-we-prefer} and \texttt{delegation} to 100\% on all four eligible victims. Refinement primarily closes coverage gaps; it cannot resolve synthesis-bound \texttt{osm-topology} or fidelity-limited \texttt{midi-protocol-lookup}, and skills solved in the first round (\texttt{protein-qc-score}, \texttt{team-composition}) are already saturated.

Holding the answered pool fixed isolates the probe-selection policy (Table~\ref{tab:paired-ablation}, Appendix~\ref{app:deployed}). Across all full-agent victims, disagreement-first selection outperforms random selection by 0.9--1.4~pp and single-shot synthesis from the same 52 answers by 1.5--5.7~pp, gaining more than 2~pp on 5--7 skills per victim against 0--3 regressions. Gains concentrate on skills where the initial clone disagrees with the victim (\texttt{demo-design-tokens} and \texttt{browser-history-acset} each gain 10~pp on Pro), while already-saturated skills tie. On Flash and Pro, five rounds of random selection merely match single-shot synthesis, showing that disagreement-guided selection drives the gain; on Luna, extra rounds also help.

\subsection{Cross-model Variations}
\label{sec:cross-model}

Figure~\ref{fig:cross-victim-results} (Appendix~\ref{app:extra}) repeats controlled reconstruction across five victims. While \texttt{stars-we-prefer} and \texttt{delegation} exceed their floors by 85 and 62~pp on every victim, other skills vary substantially by model: \texttt{lead-scoring} gains 83~pp on Kimi versus 9--11~pp on Pro and GLM, and \texttt{klingai-pricing} gains 86~pp on Pro, GLM, and Kimi versus 20~pp on Luna. Every victim leaks multiple component types, and bottlenecks are largely skill-specific rather than model-specific (e.g., \texttt{osm-topology} remains below floor on all four eligible victims at the same modular-arithmetic step). In full-agent execution (Appendix~\ref{app:deployed}), Luna applies skills reliably in OpenHands, allowing \method{} to recover 19 of 21 mined skills above floor at 82.0\% mean ASR, which shows that reconstruction succeeds through a deployed agent stack whenever the victim executes the skill.

\section{Defense Evaluation}
\label{sec:defense}

We evaluate four provider-side interventions: disclosure filters from prior work~\citep{wang2026skillstealing}, query-pattern detection, per-client query budgets, and advertisement minimization.

\paragraph{Disclosure filters enforced in the loop.} We reimplement the input intent detector (an LLM classifier flagging extraction queries) and the output filter (lexical overlap ${\ge}0.5$ with the skill body over shared spans of ${\ge}40$ characters on every parsed response), which together catch all four disclosure probes of \citet{wang2026skillstealing}. We then run \method{} end-to-end with both filters active: flagged probes are refused, high-overlap responses are withheld, and the attack proceeds using the remaining answers. We evaluate the attacker-designed probe variant, whose queries most closely resemble extraction attempts (Pro victim, Table~\ref{tab:defense}). The input detector permits nearly all probes, refusing only 4 of 5{,}508. Almost all blocking comes from the output filter, and 319 of its 398 withheld responses occur on \texttt{internal-reference}---whose benign outputs directly quote the skill text---reducing that skill's ASR to 0\%. Across the catalog, ASR remains unchanged or increases on 18 of 21 skills, showing that text-matching filters miss functional information accumulating across benign responses.

\begin{table}[t]
\centering
\small
\begin{tabular}{lrr}
\toprule
 & No filter & Filters on \\
\midrule
Queries refused (input) & -- & 4 of 5{,}508 \\
Answers withheld (output) & -- & 398 \\
Skills with non-decreasing ASR & -- & 18 of 21 \\
\texttt{internal-ref.} ASR & 26 & 0 \\
\bottomrule
\end{tabular}
\caption{End-to-end attack under enforced disclosure filters (Pro victim, attacker-designed probes, 21 skills, 5{,}508 attempted queries). Withheld responses concentrate on \texttt{internal-reference}, whose outputs quote the body; \texttt{protein-qc-score}, \texttt{team-composition}, and \texttt{working-day} remain at 100\% ASR.}
\label{tab:defense}
\end{table}

\paragraph{Query-pattern detection.} PRADA-style detectors~\citep{juuti2019prada} flag clients whose successive queries exhibit an atypical distance distribution. We fit one detector per skill on benign streams from the skill's task-input distribution at a 5\% false-positive rate (2.9\% measured on held-out benign streams) and apply it to recorded attack streams. The detector flags \method{}'s task-input streams on only 1 of 21 skills and attacker-designed probe streams on 4 of 21, where systematic sweeps such as consecutive way IDs stand out. Query-pattern detection thus constrains only attackers without realistic task inputs, who already extract less.

\paragraph{Per-client query budgets.} Most recovery occurs within the first few dozen queries. In the full-agent setting with Pro under the sequential protocol (training on every queried answer across all 21 skills), mean ASR rises from 25\% with no queries to 64\% at 42 queries, 69\% at 52, and 78\% at 82. A per-client cap tight enough to prevent reconstruction would restrict legitimate use, and an attacker can split queries across accounts.

\paragraph{Advertisement minimization.} An LLM rewrites each mined advertisement at two levels: \emph{task-only} preserves the domain and routing trigger while stripping mechanism names, constants, thresholds, table keys, and composition rules, whereas \emph{vague} retains only the broad task family. Both rewrites preserve routing accuracy and leave median ASR unchanged (Table~\ref{tab:ad-defense}). Removing cues has mixed effects: without its schema keys, \texttt{neo4j-schema} drops from 89\% to 10\%, whereas \texttt{token-cost-tracking} rises from 4\% to 54\% because the vaguer description prompts broader probes.

\begin{table}[t]
\centering
\small
\setlength{\tabcolsep}{3pt}
\begin{tabular}{lcccc}
\toprule
Description & Routing & Median $\Delta$ASR & $\downarrow{\ge}20$ & $\uparrow{\ge}20$ \\
\midrule
Original & 99.5\% & -- & -- & -- \\
Task-only & 96.2\% & 0\,pp & 3/21 & 4/21 \\
Vague & 96.2\% & 0\,pp & 2/21 & 3/21 \\
\bottomrule
\end{tabular}
\caption{Advertisement minimization: routing recall and extraction impact under rewritten descriptions. Routing is top-1 recall in closed-catalog skill selection. The last two columns count skills whose ASR decreases (helping the defender) or increases (hurting the defender) by at least 20\,pp.}
\label{tab:ad-defense}
\end{table}

\section{Conclusion}

We identify behavioral skill reconstruction as a confidentiality threat to closed-source agent skills and introduce \method{}, which turns task-valid black-box interactions into executable clones refined via disagreement signals. Across 30 skills, \method{} recovers 19 of 21 mined skills above floor on the strongest victims, while targeted interventions trace remaining failures to hypothesis quality, probe specificity, coverage, victim fidelity, recall, or synthesis. Disclosure filters, query-pattern detection, per-client budgets, and advertisement minimization offer limited protection, showing that safeguarding a skill requires limiting cumulative functional leakage over ordinary use rather than merely hiding its files.

\section*{Limitations}

\textbf{Evaluation scope.} Our metric applies to deterministic rules, thresholds, tables, formulas, and procedures; creative, interactive, and judgment-heavy skills require semantic or human evaluation.

\noindent \textbf{Model and service coverage.} We evaluate five victim models from four providers on public, open-source skills. Additional models, proprietary services, and paid marketplaces remain outside the scope of this study.


\section*{Ethical Considerations}

\paragraph{Broader impacts.}
This work identifies a confidentiality risk that arises when skill-enabled agents expose proprietary functionality through ordinary use. It helps skill developers, service providers, and platform operators evaluate leakage that artifact-focused security measures overlook.

\paragraph{Dual-use risk.}
Our attack methodology is dual-use: the probing and differential-validation strategies could reduce the effort required to imitate a deployed skill, causing economic harm to skill creators or circumventing paid access, especially for deterministic skills containing compact rules, tables, constants, or procedures.
We evaluate only publicly available, open-source skills with auditable ground truth and apply license compatibility criteria during collection. No private, paid, or commercially deployed skill is targeted. The study involves no personal or sensitive user data.

\paragraph{Use of AI assistants.} We used generative AI assistants in accordance with the ACL Policy on Publication Ethics. They assisted with literature search, prose polishing, and part of the experiment design. The authors verified all AI-assisted text, code, results, and take full responsibility for the content.

\bibliography{references}

\appendix

\renewcommand{\topfraction}{0.95}
\renewcommand{\bottomfraction}{0.8}
\renewcommand{\textfraction}{0.05}
\renewcommand{\floatpagefraction}{0.85}
\renewcommand{\dbltopfraction}{0.95}
\renewcommand{\dblfloatpagefraction}{0.85}
\setcounter{topnumber}{4}
\setcounter{dbltopnumber}{3}
\setcounter{totalnumber}{6}
\makeatletter
\setlength{\@fptop}{0pt}
\setlength{\@dblfptop}{0pt}
\makeatother

\section{Skill Types and Binding Mechanisms by Example}
\label{app:spectrum}

Table~\ref{tab:taxonomy} groups skills by \emph{component type} and names the \emph{binding mechanism} that caps each type. This appendix grounds both columns by stating the task, concealed component, and mechanism governing recovery. The advertisement exposes the task framing (e.g.\ ``detect malicious traffic'', ``compute a district bonus''); the body supplies the constants, tables, and conventions below. Table~\ref{tab:mechanisms} summarizes the six binding mechanisms with one exemplar each, followed by per-skill analyses.

\begin{table*}[t]
\centering
\small
\begin{tabular}{@{}p{2cm}p{2.2cm}p{4.6cm}p{4.6cm}@{}}
\toprule
Mechanism & Exemplar & Diagnostic (what we vary) & Signature (what it shows) \\
\midrule
Hypothesis quality & \texttt{civ6} & Supply the missing latent factor (the Government-Plaza confound) to the hypothesis. & ASR plateaus near $60\%$ until the unadvertised factor is added; once supplied, reconstruction reaches $100\%$. \\
\midrule
Probe specificity & \texttt{dapt} & Replace the attacker's coarse self-designed probes with hand-designed boundary sweeps. & ASR jumps from the $50\%$ floor to $96$--$100\%$ once exact thresholds are localized. \\
\midrule
Coverage & \texttt{codebook} & Increase the probe budget and compute a perfect-oracle upper bound. & ASR remains at $0\%$ even with a perfect oracle: 24 probes surface only 3 of 230 private entries, while held-out inputs span 127 unseen codes. \\
\midrule
Synthesis & \texttt{osm-topology} & Provide the synthesizer with the named algorithm, or apply recipe extraction. & Clones remain below floor even with faithful victim responses (\texttt{osm-topology}: $28$--$35\%$ vs.\ $38\%$ floor); supplying the algorithm or extracting the recipe lifts recovery to $100\%$. \\
\midrule
Recall & \texttt{powerlifting} & Supply the correct public formula (DOTS) directly. & ASR remains near $0\%$ with faithful victim responses ($24/24$): only the formula identity is missing, and naming DOTS resolves the gap. \\
\midrule
Victim fidelity & \texttt{r2r} & Measure the deployed victim's probe fidelity directly. & The victim answers only $12/24$ probes correctly, so noisy labels cap the clone at $0\%$; reference-oracle labels raise ASR to $100\%$. \\
\bottomrule
\end{tabular}
\caption{The six recurring mechanisms that cap held-out ASR (\S\ref{sec:results}), each with an exemplar skill, the diagnostic that isolates it, and its signature. Each mechanism is isolated by intervening on one part of the attack loop (the victim oracle, probe budget or design, or synthesizer) and checking whether ASR moves.}
\label{tab:mechanisms}
\end{table*}

\subsection{Fully Extractable: Deterministic Procedures and Small Tables}

\paragraph{drone (trajectory procedure; 100\% ASR).} The skill turns a waypoint list into a smooth quadrotor trajectory. The hidden IP is a clamped cubic-spline planner with zero velocity enforced at each waypoint and fixed per-axis acceleration limits. Ordinary trajectory requests fully exercise the procedure, so the clone reaches 100\% held-out agreement.

\paragraph{reflow (procedure conventions; 100\% ASR).} Reflow soldering attaches components by passing a circuit board through an oven with a controlled temperature profile; the solder is fully molten only above the \emph{liquidus} temperature. From thermocouple time--temperature traces the skill reports peak temperature, maximum ramp rate, and time above liquidus. The public typed interface specifies the $[100,150]^\circ$C preheat band, which we treat as public input information. The hidden conventions use linear interpolation at liquidus crossings and define the run-level peak as the \emph{minimum} of per-thermocouple maxima, with ties broken by the lowest sensor ID. Crafted traces isolate these conventions, and the clone reaches 100\% held-out agreement.


\subsection{Partially Bound: Hypothesis, Probe Specificity, and Public Data}

\paragraph{dapt (threshold rule; probe specificity).} The skill flags malicious traffic based on packet features. The IP is a conjunctive threshold rule: a port scan requires port entropy $>6.0$ bits \emph{and} SYN-only ratio $>0.7$ \emph{and} $>100$ unique ports; a DoS requires peak/average packet-rate $>20$; C2 (command-and-control) beaconing requires inter-arrival-time CV $<0.5$. Hand-designed boundary sweeps localize every cutoff (96--100\% ASR). The main protocol's coarse probes miss the exact constants and keep ASR at the 50\% balanced-class floor, and fully autonomous probes reach only 61--75\% (Table~\ref{tab:autonomy}), showing that probe specificity is the bottleneck.

\paragraph{civ6 (rule table + confound; hypothesis quality).} The skill totals Civilization VI district adjacency bonuses. Beyond a per-district bonus table ($+2/{+}1/{+}0.5$ terms per neighbor type), the IP includes a latent \emph{confound}: an adjacent Government Plaza adds $+1$ to specialty districts, a factor that the advertisement never names. The attacker recovers the explicit table but frequently omits this unadvertised factor, so ASR plateaus near 60\%.

\paragraph{labunit (lookup/conversion data; public data).} The skill harmonizes clinical lab values to standard units using a table of $\sim$60 analytes (substances measured by lab tests) with valid ranges and conversion factors, applied via a range-triggered rule. Enumerating analytes recovers most of the table (77\% ASR), with standard clinical knowledge supplying part of the conversions.

\subsection{Hard-bound: Victim Fidelity, Coverage, Synthesis, and Recall}

\paragraph{r2r (numeric linearization; victim fidelity).} Roll-to-roll (R2R) manufacturing moves a flexible material web between rollers, as in printed-electronics or film production. The skill linearizes the web's tension--velocity dynamics into a discrete state-space $(A,B)$ for model-predictive control, where Jacobian entries depend on hidden plant constants (modulus--area product $EA$, roller radius $R$, inertia $J$, web length $L$). The deployed victim computes this correctly on only 12/24 probes, so the attacker trains on noisy labels and scores 0\% on the constant-bearing entries. Replacing victim responses with reference-oracle labels raises ASR to 100\%, isolating victim fidelity as the bottleneck.

\paragraph{codebook (private table; coverage).} The skill normalizes free-text failure reasons to standard manufacturing codes via per-product codebooks ($230$ code$\rightarrow$label entries: $72{+}80{+}78$) plus a weighted-matching procedure. A 24-probe budget surfaces only 3 of the 230 entries (0\% ASR). Held-out inputs span 127 unseen codes, so even a perfect oracle caps recovery at the $\sim$2\% of held-out mass carried by those three codes; probe coverage sets the ceiling.

\paragraph{osm-topology (derived procedure; synthesis).} \texttt{osm-topology} (OpenStreetMap) applies a modular hue$\rightarrow$ternary-color assignment (coloring each feature into one of three classes) whose exact arithmetic the clone fails to express; closed-loop ASR remains at 28--35\% (below its 38\% floor), and even a 64k-token synthesis budget yields no valid program. Stateful procedures alone do not block synthesis: \texttt{cache} replays a KV-cache trace under the S3-FIFO policy \citep{yang2023s3fifo} (small/main/ghost FIFO queues, a $0$--$3$ frequency counter, second-chance admission, and longest-prefix hit semantics), yet clones reach 99.8--100\% ASR on Flash, Pro, GLM, and Kimi and 58\% on Luna (floor 18\%).

\paragraph{powerlifting (public formula; recall).} The skill computes a normalized competition score from a lifter's sex, bodyweight, and total by multiplying the total by a bodyweight polynomial. Powerlifting federations use several published formulas: Wilks \citep{vanderburgh1999wilks}, IPF GL \citep{maiwald2018ipfreview,ipf2020gl}, Glossbrenner, and DOTS \citep{powerliftprodots}. The skill body specifies DOTS: total $\times$ a sex-specific quartic coefficient $500/\mathrm{poly}_4(\mathrm{BW})$, with bodyweight clamped to $[40,210]\,\mathrm{kg}$ (men) and $[40,150]\,\mathrm{kg}$ (women). Because the coefficients are \emph{public} and the advertisement names no formula, the only hidden information is the \emph{choice} of DOTS, reducing reconstruction to recalling the right formula. A wrong choice leaves ASR near $0\%$ even with faithful victim responses ($24/24$): the attacker defaults to the best-known formula, Wilks, and never proposes DOTS. Because Wilks and DOTS agree within $\sim$1\% at mid-range bodyweights ($60$--$125$\,kg), a Wilks clone passes agreement-based validation; although they diverge by $5$--$6\%$ at 40 and 200\,kg, data can only discriminate among hypotheses the attacker proposes. Supplying the name ``DOTS'' resolves the gap completely.

\section{Benchmark Construction Details}
\label{app:benchmark}

This appendix expands the selection pipeline and test construction summarized in \S\ref{sec:evaluation}.

\subsection{Mining and Licensing}

The 9 code-execution skills come from SkillsBench \citep{li2026skillsbench}, which curates skills that bundle scripts and deterministic verifiers. The 21 mined data/rule skills come from two crawls: the SkillRet corpus of $17{,}810$ public skills curated from $22{,}795$ marketplace listings \citep{cho2026skillret}, and a broader registry crawl spanning hundreds of source repositories. We retain MIT/Apache-or-equivalent licensed packages and deduplicate by source URL, repository, and skill name. The deterministic-ground-truth and marginal-functionality filters produce the purpose-selected evaluation suite described in \S\ref{sec:suite}.

\subsection{Selection Requirements}

\paragraph{Observable marginal functionality.} A target must be IP-positive for at least one victim model: the model applies the body reliably and cannot reproduce the behavior from the advertisement alone. For data/rule skills, the controlled comparison requires $\mathrm{fid}(A_{m,s})\ge 0.6$ and a fidelity gain over the advertisement baseline of at least $0.25$. The paired full-agent comparison asks the same question after skill selection, file access, tool use, and execution are delegated to OpenHands. Table~\ref{tab:ipcounts} summarizes the resulting per-victim IP-positive counts.

\paragraph{Exact ground truth.} A target must expose a deterministic function that we can call or reimplement to label arbitrary held-out inputs, restricting the benchmark to skills amenable to exact-equivalence evaluation.

\begin{table}[t]
\centering
\small
\setlength{\tabcolsep}{4pt}
\begin{tabular}{lccc}
\toprule
Victim & Code (9) & Mined (21) & All (30) \\
\midrule
DeepSeek-V4-Pro & 9 & 20 & 29 \\
DeepSeek-V4-Flash & 9 & 14 & 23 \\
Kimi-K2.6 & 9 & 13 & 22 \\
GLM-5.1 & 9 & 12 & 21 \\
GPT-5.6-Luna & 2 & 13 & 15 \\
\bottomrule
\end{tabular}
\caption{IP-positive cells in the curated 30-skill suite. The 9 SkillsBench targets are confirmed through the full-agent pipeline; the 21 mined targets use the controlled body-versus-advertisement comparison. Counts define the cells entering controlled extraction analysis.}
\label{tab:ipcounts}
\end{table}

\subsection{Oracles, Discriminating Inputs, and Splits}

For each surviving skill, we author an adapter exposing a deterministic reference oracle $f_s^\star$ reimplemented from the private body. The oracle captures the \emph{functional core} of the body (the deterministic mapping from typed inputs to outputs) while excluding the public advertisement, instructional framing, worked examples, and formatting. The adapter also provides a typed input/output schema and an input generator emphasizing \emph{discriminating} inputs $D(c)$ on which correct and incorrect implementations diverge. Two skills receive explicit rebalancing: \texttt{dapt} uses a class-balanced held-out set (50\% malicious rate and 50\% trivial floor, versus an 89\% deployment base rate), and \texttt{r2r} scores only the Jacobian entries encoding private plant constants (giving all-zero and identity clones a 0\% floor). Test inputs are disjoint from probe and validation inputs wherever the input domain permits. Mined-skill clones use $n{=}150$ held-out test inputs, and SkillsBench clones use $n{=}800$. The mined controlled protocol uses 52 acquisition probes and 30 validation queries; SkillsBench attacks use $K{=}24$ victim probes.

\section{Additional Results}
\label{app:extra}

This appendix supplements \S\ref{sec:results} with per-skill matrices, attacker-model ablations, refinement breakdowns, and sampling-stability diagnostics.

\paragraph{API reasoning settings.} We use the API defaults at experiment time unless specified. DeepSeek-V4-Flash (Preview) and Pro \citep{deepseekai2026deepseekv4} inherit \texttt{thinking=enabled} and \texttt{reasoning\_effort=high}. GLM-5.1 \citep{glm2026} uses \texttt{thinking=disabled}. GPT-5.6-Luna receives the OpenAI reasoning endpoint's default effort. Our controlled Kimi-K2.6 requests omitted the \texttt{thinking} field, following Moonshot's documented API usage for thinking mode \citep{moonshot2026kimik26}.

\subsection{Controlled Cross-model Results}

\begin{figure*}[t]
\centering
\includegraphics[width=\textwidth]{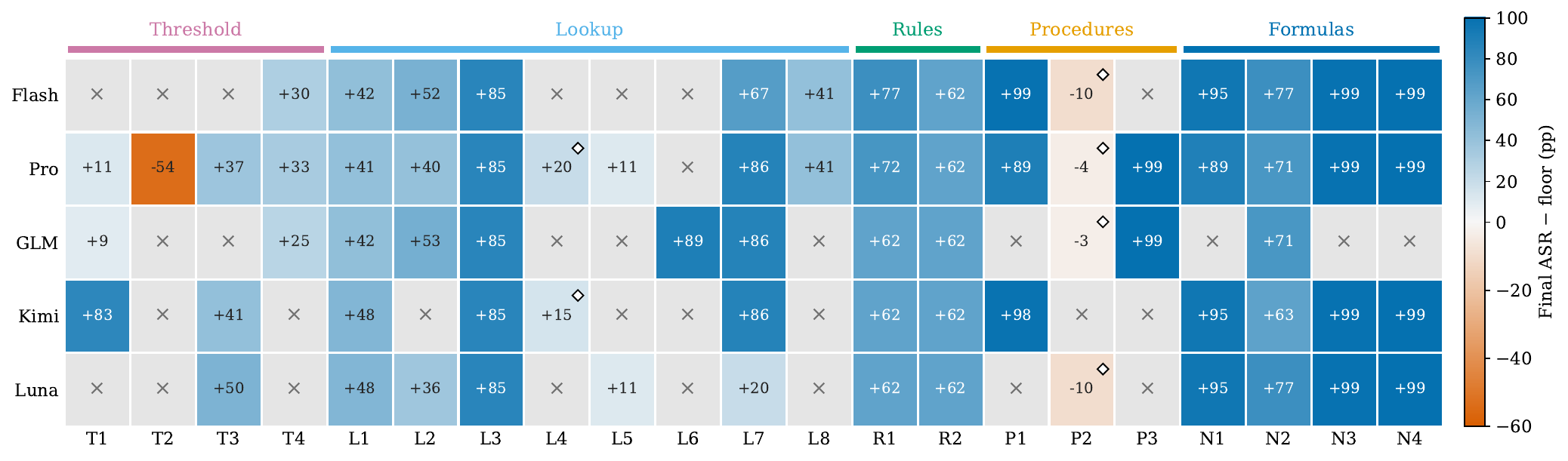}
\caption{Cross-victim reconstruction results. Skills grouped as threshold (T), lookup (L), rules (R), procedure (P), and numeric-formula (N), with codes mapped in Table~\ref{tab:heatmap-codes}. Cells report ASR minus per-skill floor; gray $\times$ cells lack measurable marginal functionality, and open diamonds mark synthesis-bound outcomes.}
\label{fig:cross-victim-results}
\end{figure*}

Table~\ref{tab:mined} provides the controlled matrix summarized in \S\ref{sec:results}, where direct body placement measures reconstruction conditional on successful skill delivery across all 21 mined targets.

Table~\ref{tab:heatmap-codes} maps the compact skill codes in Figure~\ref{fig:cross-victim-results} to their full names.

\begin{table}[t]
\centering
\small
\renewcommand{\arraystretch}{0.9}
\setlength{\tabcolsep}{4pt}
\begin{tabular}{@{}lp{0.80\columnwidth}@{}}
\toprule
Code & Skill \\
\midrule
T1 & \path{lead-scoring} \\
T2 & \path{check-secrets} \\
T3 & \path{security-environment-standards} \\
T4 & \path{rbac-validator} \\
\addlinespace[1pt]
L1 & \path{internal-reference} \\
L2 & \path{demo-design-tokens} \\
L3 & \path{stars-we-prefer} \\
L4 & \path{browser-history-acset} \\
L5 & \path{neo4j-schema} \\
L6 & \path{isocom} \\
L7 & \path{klingai-pricing} \\
L8 & \path{midi-protocol-lookup} \\
\addlinespace[1pt]
R1 & \path{bellog-structure} \\
R2 & \path{delegation} \\
\addlinespace[1pt]
P1 & \path{token-cost-tracking} \\
P2 & \path{osm-topology} \\
P3 & \path{working-day} \\
\addlinespace[1pt]
N1 & \path{labor-rate} \\
N2 & \path{knowledge-worker-salaries} \\
N3 & \path{protein-qc-score} \\
N4 & \path{team-composition} \\
\bottomrule
\end{tabular}
\caption{Codebook for Figure~\ref{fig:cross-victim-results}. Prefixes T/L/R/P/N denote threshold, lookup, rule-composition, procedure, and numeric-formula components.}
\label{tab:heatmap-codes}
\end{table}

\begin{table}[t]
\centering
\small
\resizebox{\columnwidth}{!}{%
\begin{tabular}{lrrrrrr}
\toprule
 & & \multicolumn{5}{c}{controlled closed-loop ASR (\%)} \\
\cmidrule(lr){3-7}
Skill & floor & F & P & G & K & L \\
\midrule
stars-we-prefer & 15 & 100 & 100 & 100 & 100 & 100 \\
team-composition & 1 & 100 & 100 & -- & 100 & 100 \\
token-cost-tracking & 1 & 100 & 91 & -- & 99 & -- \\
working-day & 1 & -- & 100 & 100 & -- & -- \\
delegation & 38 & 100 & 100 & 100 & 100 & 100 \\
rbac-validator & 66 & 96 & 99 & 91 & -- & -- \\
security-env-standards & 50 & -- & 87 & -- & 91 & 100 \\
neo4j-schema & 81 & -- & 92 & -- & -- & 92 \\
klingai-pricing & 14 & 81 & 100 & 100 & 100 & 34 \\
check-secrets & 65 & -- & 11 & -- & -- & -- \\
bellog-structure & 11 & 88 & 83 & 73 & 73 & 73 \\
knowledge-worker-salaries & 9 & 86 & 81 & 81 & 72 & 86 \\
internal-reference & 27 & 69 & 68 & 69 & 75 & 75 \\
isocom & 11 & -- & -- & 100 & -- & -- \\
demo-design-tokens & 6 & 58 & 46 & 59 & -- & 42 \\
midi-protocol-lookup & 9 & 50 & 50 & -- & -- & -- \\
labor-rate & 5 & 100 & 94 & -- & 100 & 100 \\
osm-topology & 38 & \textit{28} & \textit{34} & \textit{35} & -- & \textit{28} \\
lead-scoring & 17 & -- & 28 & 25 & 100 & -- \\
protein-qc-score & 1 & 100 & 100 & -- & 100 & 100 \\
browser-history-acset & 31 & -- & \textit{51} & -- & \textit{45} & -- \\
\bottomrule
\end{tabular}%
}
\caption{Controlled closed-loop ASR (\%, $n{=}150$) on the 21 mined data/rule skills with a Flash attacker. F/P/G/K/L denote victim models (Flash, Pro, GLM-5.1, Kimi-K2.6, GPT-5.6-Luna); ``--'' marks cells without measurable marginal functionality in the controlled setting. Italic values are synthesis-bound. The matrix measures reconstruction conditional on successful skill delivery.}
\label{tab:mined}
\end{table}

\subsection{Attacker Model}
\label{app:attacker-model}

Table~\ref{tab:autonomy} compares the primary Flash attacker against the stronger Pro attacker under the fully autonomous \method{} protocol (LLM-designed probes, LLM clone induction) on the five code-execution skills, with Pro as the victim throughout.

\begin{table}[t]
\centering
\small
\begin{tabular}{llcc}
\toprule
 & & Pro & Flash \\
Skill & Type & attacker & attacker \\
\midrule
dapt & threshold & 74.8\% & 60.6\% \\
civ6 & rule & 55.1\% & 43.5\% \\
labunit & data & 84.9\% & 13.2\% \\
reflow & procedure & 100.0\% & 0.0\% \\
cache & stateful & 14.0\% & 57.2\% \\
\bottomrule
\end{tabular}
\caption{Held-out ASR (\%) by attacker model under the fully autonomous \method{} protocol (LLM-designed probes, LLM clone induction) on the five code-execution SkillsBench targets (Pro victim throughout). Each cell is a single induction without validation selection.}
\label{tab:autonomy}
\end{table}

Pro outperforms Flash on four of five skills: by 14.2~pp on \texttt{dapt}, 11.6~pp on \texttt{civ6}, 71.7~pp on \texttt{labunit}, and 100~pp on \texttt{reflow}, where every Flash attempt exhausts the 64k-token output budget before emitting a clone. Flash leads on \texttt{cache} (57.2\% vs.\ 14.0\%). Because each attacker designs its own probes and induces the clone in a single pass, synthesis errors go uncorrected. In the main SkillsBench protocol (\S\ref{sec:results-bottlenecks}), which synthesizes three candidates and selects by validation agreement, Flash reaches 100\% on \texttt{reflow}, 74.3\% on \texttt{labunit}, and 99.8\% on \texttt{cache} against the same Pro victim. Synthesis averages approximately \$0.10 per Flash attack and \$0.14 per Pro attack: although Pro's output tokens cost three times as much, Flash's truncated attempts emit twice as many. We retain Flash as the primary attacker because it is more economical, provides a conservative lower bound on what stronger attackers can recover, and exposes synthesis bottlenecks that Pro partially masks.

\subsection{Closed-loop Refinement}
\label{app:close-loop-refine}

Table~\ref{tab:cl-all} reports per-skill closed-loop lift by component type across four victim models (Flash attacker). Each round adds the next $10$ task inputs ($k_0{=}12$, $4$ rounds), and the best round is selected by victim-only validation agreement. This protocol synthesizes one candidate per round from a 12k-token starting budget, extended if the response is truncated. Cells that are not IP-positive for a given victim are excluded from that victim's column (Table~\ref{tab:ipcounts}). As summarized in \S\ref{sec:closed-loop}, closed-loop refinement adds $+25$~pp (Pro), $+28$~pp (GLM), $+30$~pp (Kimi), and $+27$~pp (Luna) across the 13 skills with mean lift ${>}5$~pp. Averaging across all IP-positive cells instead gives $+17.0$~pp (Pro, $n{=}20$), $+18.7$~pp (GLM, $n{=}12$), $+25.5$~pp (Kimi, $n{=}13$), and $+20.5$~pp (Luna, $n{=}13$), since cells already at floor or ceiling before refinement dilute the mean.

\begin{table}[htbp]
\centering
\scriptsize
\setlength{\tabcolsep}{2.2pt}
\begin{tabular}{lrrrrrrrrr}
\toprule
 & & \multicolumn{2}{c}{Pro} & \multicolumn{2}{c}{GLM} & \multicolumn{2}{c}{Kimi} & \multicolumn{2}{c}{Luna} \\
\cmidrule(lr){3-4} \cmidrule(lr){5-6} \cmidrule(lr){7-8} \cmidrule(lr){9-10}
Skill & Floor & CL & $\Delta$ & CL & $\Delta$ & CL & $\Delta$ & CL & $\Delta$ \\
\midrule
\multicolumn{10}{l}{\emph{Threshold / decision rules (4 skills)}} \\
lead-scoring       & 17 & 28 & +7 & 25 & +5 & 100 & +77 & -- & -- \\
check-secrets      & 65 & 11 & +0 & -- & -- & -- & -- & -- & -- \\
sec-env-std        & 50 & 87 & +14 & -- & -- & 91 & +13 & 100 & +4 \\
rbac-validator     & 66 & 99 & +7 & 91 & -3 & -- & -- & -- & -- \\
\textit{Group avg.} & 49 & 56 & +7 & 58 & +1 & 95 & +45 & 100 & +4 \\
\addlinespace[2pt]
\multicolumn{10}{l}{\emph{Lookup tables / data (8 skills)}} \\
internal-ref       & 27 & 68 & +49 & 69 & +47 & 75 & +52 & 75 & +50 \\
design-tokens      & 6 & 46 & +31 & 59 & +29 & -- & -- & 42 & +15 \\
stars-we-prefer    & 15 & 100 & +10 & 100 & +10 & 100 & +10 & 100 & +21 \\
browser-hist       & 31 & 51 & +0 & -- & -- & 45 & +21 & -- & -- \\
neo4j-schema       & 81 & 92 & +0 & -- & -- & -- & -- & 92 & +11 \\
isocom             & 11 & -- & -- & 100 & +0 & -- & -- & -- & -- \\
klingai-pricing    & 14 & 100 & +16 & 100 & +16 & 100 & +16 & 34 & -11 \\
midi-protocol      & 9 & 50 & +0 & -- & -- & -- & -- & -- & -- \\
\textit{Group avg.} & 24 & 72 & +15 & 86 & +20 & 80 & +25 & 69 & +17 \\
\addlinespace[2pt]
\multicolumn{10}{l}{\emph{Rule composition (2 skills)}} \\
bellog             & 11 & 83 & +37 & 73 & +32 & 73 & +27 & 73 & +27 \\
delegation         & 38 & 100 & +31 & 100 & +31 & 100 & +31 & 100 & +31 \\
\textit{Group avg.} & 25 & 92 & +34 & 87 & +31 & 87 & +29 & 87 & +29 \\
\addlinespace[2pt]
\multicolumn{10}{l}{\emph{Procedures / algorithms (3 skills)}} \\
token-cost         & 1 & 91 & +13 & -- & -- & 99 & +0 & -- & -- \\
osm-topology       & 38 & 34 & +6 & 35 & +7 & -- & -- & 28 & +0 \\
working-day        & 1 & 100 & +0 & 100 & +0 & -- & -- & -- & -- \\
\textit{Group avg.} & 13 & 75 & +6 & 67 & +3 & 99 & +0 & 28 & +0 \\
\addlinespace[2pt]
\multicolumn{10}{l}{\emph{Numeric formulas (4 skills)}} \\
labor-rate         & 5 & 94 & +70 & -- & -- & 100 & +38 & 100 & +63 \\
knowl-worker-sal   & 9 & 81 & +51 & 81 & +51 & 72 & +48 & 86 & +56 \\
protein-qc         & 1 & 100 & +0 & -- & -- & 100 & +0 & 100 & +0 \\
team-composition   & 1 & 100 & +0 & -- & -- & 100 & +0 & 100 & +0 \\
\textit{Group avg.} & 4 & 94 & +30 & 81 & +51 & 93 & +22 & 96 & +30 \\
\bottomrule
\end{tabular}
\caption{Closed-loop refinement across all 21 mined targets grouped by component type (Flash attacker). CL = closed-loop ASR (\%); $\Delta$ = lift over single-shot (pp). Group averages are unweighted across IP-positive cells in each group; ``--'' marks cells that are not IP-positive (Table~\ref{tab:mined}).}
\label{tab:cl-all}
\end{table}

\subsection{Recipe Extraction}
\label{app:recipe-extraction}

Recipe extraction asks the agent to solve benign tasks while showing step-by-step reasoning. It targets synthesis-bound procedures, complementing the closed loop's coverage gains. Table~\ref{tab:recipe} reports Pro and Luna victims with a Flash attacker and majority-class fallback disabled. Recipe extraction resolves \texttt{osm-topology}'s synthesis bottleneck for both victims (Pro $32.5{\to}100\%$, Luna $0{\to}100\%$) and leaves coverage-limited \texttt{neo4j-schema} and already-extractable skills unchanged. We omit recipe extraction from the main attack because it helps only on synthesis-bound skills and explicitly requesting step-by-step reasoning may alert defenders.

\begin{table*}[htbp]
\centering
\small
\setlength{\tabcolsep}{6pt}
\begin{tabular}{lrrrrrr}
\toprule
 & \multicolumn{3}{c}{Pro victim} & \multicolumn{3}{c}{Luna victim} \\
\cmidrule(lr){2-4} \cmidrule(lr){5-7}
Skill & I/O & +Recipe & $\Delta$ & I/O & +Recipe & $\Delta$ \\
\midrule
osm-topology & 32.5 & 100.0 & +67.5 & 0.0 & 100.0 & +100.0 \\
neo4j-schema & 20.0 & 25.0 & +5.0 & 20.0 & 25.0 & +5.0 \\
knowledge-worker-salaries & 28.7 & 28.7 & +0.0 & 31.2 & 31.2 & +0.0 \\
security-env-standards & 83.8 & 100.0 & +16.2 & 83.8 & 100.0 & +16.2 \\
\bottomrule
\end{tabular}
\caption{Recipe extraction transfers across victim model families: step-by-step reasoning resolves the synthesis bottleneck that caps \texttt{osm-topology} at 0\% I/O ASR on Luna (reaching 100\%), matching Pro. Recipe lift is consistent across both victims on the other three skills.}
\label{tab:recipe}
\end{table*}

\subsection{Query Contribution: Per-skill Results}
\label{app:zeroquery}

This appendix gives the per-skill results behind \S\ref{sec:queries} (Pro victim, controlled setting, Flash attacker, $n{=}150$ held-out inputs). Table~\ref{tab:zeroquery} lists each skill's floor, its zero-query ASR (clone written from the advertisement and task interface alone, mean of five runs), the plain-LLM baseline, and \method{}. With no queries, 13 of 21 skills stay at or below their floor; the skills that score high without queries implement public conventions or price sheets, such as \texttt{working-day} and \texttt{token-cost-tracking}.

\begin{table}[htbp]
\centering
\footnotesize
\setlength{\tabcolsep}{3pt}
\begin{tabular}{lrrrr}
\toprule
Skill & Floor & No query & Plain & Ours \\
\midrule
\multicolumn{5}{l}{\emph{Rule composition}} \\
bellog & 2 & 0 & 31 & 100 \\
delegation & 36 & 0 & 47 & 100 \\
\multicolumn{5}{l}{\emph{Numeric formulas}} \\
knowledge-worker-sal. & 9 & 0 & 6 & 86 \\
labor-rate & 3 & 0 & 1 & 88 \\
protein-qc & 1 & 6 & 100 & 100 \\
team-composition & 1 & 0 & 100 & 100 \\
\multicolumn{5}{l}{\emph{Lookup tables}} \\
browser-history & 31 & 17 & 31 & 31 \\
design-tokens & 7 & 14 & 0 & 73 \\
internal-ref. & 26 & 0 & 26 & 75 \\
isocom & 17 & 16 & 13 & 47 \\
klingai-pricing & 15 & 55 & 7 & 100 \\
midi-protocol & 8 & 12 & 7 & 61 \\
neo4j-schema & 55 & 36 & 45 & 55 \\
stars-we-prefer & 14 & 0 & 14 & 100 \\
\multicolumn{5}{l}{\emph{Threshold rules}} \\
check-secrets & 67 & 15 & 9 & 41 \\
lead-scoring & 4 & 0 & 0 & 7 \\
rbac-validator & 70 & 80 & 91 & 93 \\
sec-env-standards & 51 & 76 & 49 & 95 \\
\multicolumn{5}{l}{\emph{Procedures}} \\
osm-topology & 39 & 37 & 31 & 31 \\
token-cost & 1 & 82 & 100 & 100 \\
working-day & 1 & 86 & 100 & 100 \\
\bottomrule
\end{tabular}
\caption{Per-skill query contribution (ASR \%, Pro victim). No query: clone written from the advertisement alone, mean of five runs. Plain: plain-LLM baseline. Ours: \method{}.}
\label{tab:zeroquery}
\end{table}

\paragraph{Altered skills.} Table~\ref{tab:altered} reports the altered skills, whose hidden constants and rules we change in both the skill body and the oracle. \method{} exceeds the zero-query baseline on 10 of 11; on \texttt{isocom}, neither recovers the altered table.

\begin{table}[htbp]
\centering
\footnotesize
\setlength{\tabcolsep}{3pt}
\begin{tabular}{lrr}
\toprule
Altered skill & Zero-query & \method{} \\
\midrule
labor-rate & 0 & 23 \\
team-composition & 0 & 100 \\
bellog & 0 & 87 \\
knowledge-worker-sal. & 2 & 86 \\
protein-qc & 16 & 100 \\
isocom & 20 & 0 \\
klingai-pricing & 32 & 61 \\
rbac-validator & 59 & 90 \\
working-day & 62 & 99 \\
sec-env-standards & 64 & 77 \\
token-cost & 75 & 100 \\
\bottomrule
\end{tabular}
\caption{Altered skills (ASR \%, Pro victim). Zero-query is the mean of five runs.}
\label{tab:altered}
\end{table}

\paragraph{Rule-level recovery.} For every skill with at least two exercised paths, Table~\ref{tab:paths} reports test-set statement coverage of the reference oracle, the number of exercised execution paths (a path is the set of reference statements an input executes), the count of exact paths, and per-path accuracy (clone accuracy averaged over paths) alongside ASR.

\begin{table}[htbp]
\centering
\footnotesize
\setlength{\tabcolsep}{2pt}
\begin{tabular}{lrrrrr}
\toprule
Skill & Stmt. & Paths & Exact & Path acc. & ASR \\
\midrule
bellog & 73 & 6 & 6 & 100.0 & 100.0 \\
check-secrets & 100 & 8 & 3 & 44.2 & 40.7 \\
design-tokens & 83 & 2 & 0 & 42.2 & 72.7 \\
internal-ref. & 100 & 2 & 1 & 82.9 & 74.7 \\
isocom & 91 & 4 & 1 & 50.2 & 46.7 \\
klingai-pricing & 75 & 2 & 2 & 100.0 & 100.0 \\
knowledge-worker & 93 & 5 & 3 & 78.4 & 86.0 \\
labor-rate & 100 & 2 & 1 & 82.7 & 88.0 \\
neo4j-schema & 92 & 4 & 1 & 33.6 & 54.7 \\
osm-topology & 100 & 3 & 1 & 33.3 & 31.3 \\
rbac-validator & 96 & 24 & 16 & 90.6 & 92.7 \\
working-day & 100 & 2 & 2 & 100.0 & 100.0 \\
\bottomrule
\end{tabular}
\caption{Test adequacy and path-level accuracy (\%, Pro victim, \method{} clones). Stmt.: share of the reference's statements the test inputs execute. Path acc.: clone accuracy averaged over paths.}
\label{tab:paths}
\end{table}

\subsection{Robustness to Victim Sampling}
\label{app:temp}

We test whether recovery depends on victim sampling by requesting temperatures $\{0,0.6,1.0\}$ covering the vendors' recommended settings \citep{deepseek2026params}. GLM-5.1 runs with thinking disabled, so these represent three distinct temperatures. Flash and Pro run in thinking mode, which ignores the temperature argument \citep{deepseek2026thinking}; for them, the three settings correspond to repeated runs under default sampling, measuring run-to-run stability. We evaluated Kimi-K2.6 only at $T{=}1$ in thinking mode \citep{moonshot2026kimik26}, and Luna's reasoning endpoint exposes no temperature parameter, so both are omitted from the sweep. The Flash attacker samples at DeepSeek's defaults throughout. For each setting after the first, we draw three seeds for the body-versus-advertisement comparison and two for extraction, reported relative to the first setting.

Marginal functionality remains stable (Table~\ref{tab:temp-function}): 93--100\% of IP-positive cells from the first setting retain that status across a majority of seeds, and the mean fidelity gain over the advertisement remains at least $+0.6$ in the final setting. Extraction is similarly stable (Table~\ref{tab:temp-attack}). On GLM---the one victim whose temperature parameter takes effect---single-observation ASR shifts only from 62\% to 59\% and 59\%, confirming that raising temperature does not protect the skill. On Flash and Pro, repeated runs vary by at most 13~pp in pooled ASR (Pro: 67\%, 54\%, 66\%), well within the $\pm20$--$41$ spread across skills.

\begin{table}[htbp]
\centering
\scriptsize
\setlength{\tabcolsep}{2.5pt}
\resizebox{\columnwidth}{!}{%
\begin{tabular}{lcccc}
\toprule
Victim & $T{=}0$ & $T{=}0.6$ & $T{=}1.0$ & Margin @ $T{=}1$ \\
\midrule
DeepSeek-Flash & 14 & 14/14 (100\%) & 13/14 (93\%) & $+0.75$ \\
DeepSeek-Pro   & 17 & 16/17 (94\%)  & 17/17 (100\%) & $+0.63$ \\
GLM-5.1        & 10 & 10/10 (100\%) & 10/10 (100\%) & $+0.62$ \\
\bottomrule
\end{tabular}%
}
\caption{Stability of controlled marginal-functionality classification across requested victim temperatures ($n{=}10$, $3$ seeds after the first setting). Retention is relative to each victim's $T{=}0$ set. DeepSeek's thinking mode ignores temperature, so the Flash and Pro columns are repeated runs at default sampling; only GLM's are real temperatures. We omit Kimi because we evaluated it only at $T{=}1$.}
\label{tab:temp-function}
\end{table}

\begin{table}[htbp]
\centering
\scriptsize
\begin{tabular}{llccc}
\toprule
victim & cond & $T{=}0$ & $T{=}0.6$ & $T{=}1.0$ \\
\midrule
DeepSeek-Flash & single & $72\pm22$ & $73\pm20$ & $73\pm28$ \\
               & mv     & $60\pm40$ & $68\pm27$ & $66\pm32$ \\
DeepSeek-Pro   & single & $67\pm27$ & $54\pm41$ & $66\pm33$ \\
               & mv     & $67\pm32$ & $58\pm34$ & $63\pm36$ \\
GLM-5.1        & single & $62\pm26$ & $59\pm29$ & $59\pm34$ \\
               & mv     & $59\pm34$ & $52\pm32$ & $56\pm31$ \\
Kimi-K2.6$^\dagger$ & single & -- & -- & $57\pm34$ \\
               & mv     & -- & -- & $60\pm38$ \\
\bottomrule
\end{tabular}
\caption{Extraction ASR (\%, $\pm$ stdev) vs.\ requested victim temperature $T$ (Flash attacker, $8$ skills, $K{=}12$, $n{=}150$). single: one observation per probe; mv: modal (majority) answer over $3$ repeats. Flash and Pro ignore $T$ in thinking mode, so their columns are repeated runs; GLM's are real temperatures. $^\dagger$We evaluated Kimi only at $T{=}1$.}
\label{tab:temp-attack}
\end{table}

\subsection{Probe-budget Sensitivity}
\label{app:probe-ablation}

We vary the initial probe budget $k_0\in\{6,12,24,48,96\}$ for eight skills spanning four bottleneck types and measure single-shot ASR with a Flash attacker and Pro and Luna victims (rounds${=}0$, i.e., no closed loop; otherwise identical protocol to \S\ref{sec:evaluation}). Table~\ref{tab:probe-ablation} reports the full sweep: only a subset of skills improves steadily as $k_0$ increases.

\begin{table}[htbp]
\centering
\scriptsize
\setlength{\tabcolsep}{1pt}
\begin{tabular}{llccccc}
\toprule
Skill & Victim & $k_0{=}6$ & 12 & 24 & 48 & 96 \\
\midrule
\texttt{internal-reference} & Pro & .093 & .000 & .520 & .687 & .820 \\
 & Luna & .093 & .220 & .540 & .740 & .887 \\
\texttt{klingai-pricing} & Pro & .640 & .820 & 1.00 & 1.00 & 1.00 \\
 & Luna & .380 & .253 & .253 & 1.00 & .253 \\
\texttt{midi-protocol-lookup} & Pro & .393 & .380 & .053 & .520 & .607 \\
 & Luna & .480 & .520 & .527 & .047 & .600 \\
\texttt{demo-design-tokens} & Pro & .027 & .000 & .233 & .340 & .740 \\
 & Luna & .320 & .153 & .487 & .033 & .053 \\
\texttt{neo4j-schema} & Pro & .793 & .933 & .800 & .807 & .820 \\
 & Luna & .793 & .920 & .800 & .807 & .820 \\
\texttt{stars-we-prefer} & Pro & .573 & .893 & 1.00 & 1.00 & 1.00 \\
 & Luna & .573 & .760 & 1.00 & 1.00 & 1.00 \\
\texttt{bellog-structure} & Pro & .073 & .573 & .787 & .787 & 1.00 \\
 & Luna & .247 & .500 & .680 & .093 & .700 \\
\texttt{osm-topology} & Pro & .193 & .207 & .313 & .353 & .320 \\
 & Luna & .193 & .207 & .227 & .267 & .360 \\
\bottomrule
\end{tabular}%
\caption{Single-shot ASR with increasing probe budgets $k_0$ (Flash attacker, rounds${=}0$, no closed loop). With Pro, \texttt{internal-reference}, \texttt{klingai-pricing}, \texttt{stars-we-prefer}, \texttt{bellog-structure}, and \texttt{demo-design-tokens} rise with $k_0$, reflecting a coverage bottleneck. With Luna, \texttt{klingai-pricing}, \texttt{bellog-structure}, and \texttt{demo-design-tokens} fluctuate across budgets, so coverage sensitivity is model-conditional. \texttt{neo4j-schema} and \texttt{osm-topology} stay flat due to a high majority-class floor and synthesis limits, respectively.}
\label{tab:probe-ablation}
\end{table}

\paragraph{Single-shot versus adaptive refinement.} To separate disagreement-driven selection from the number of synthesis rounds, we add a single-shot arm to the paired full-agent protocol (Pro victim) that synthesizes once---using the same three candidates and validation selection---from the exact 52 answers accumulated by the fixed policy and is scored on the same test inputs. Table~\ref{tab:same-budget} compares the three arms: single-shot averages 80.2\%, the fixed policy 80.8\%, and the adaptive policy 82.1\%, with the adaptive policy outperforming the fixed policy by more than 2~pp on 6 skills and regressing on none. On Flash and Luna (20 skills each, excluding \texttt{osm-topology}), the corresponding means are 88.3\%, 88.2\%, and 89.8\% (Flash; 5 improved, 3 regressed) and 78.9\%, 83.6\%, and 84.6\% (Luna; 7 improved, 1 regressed).

\begin{table}[htbp]
\centering
\small
\setlength{\tabcolsep}{3pt}
\begin{tabular}{lrrrr}
\toprule
Skill & Floor & Single-shot & Fixed & Adaptive \\
\midrule
browser-hist & 31 & 90 & 90 & 100 \\
design-tokens & 7 & 76 & 77 & 86 \\
sec-env-std$^*$ & 51 & 92 & 92 & 100 \\
osm-topology & 39 & 24 & 37 & 31 \\
internal-ref & 26 & 82 & 82 & 85 \\
rbac-validator & 70 & 87 & 88 & 89 \\
lead-scoring & 17 & 31 & 31 & 31 \\
protein-qc & 1 & 100 & 100 & 100 \\
stars-we-prefer & 14 & 100 & 100 & 100 \\
check-secrets & 67 & 84 & 79 & 84 \\
token-cost & 1 & 100 & 100 & 100 \\
isocom & 11 & 100 & 100 & 100 \\
klingai-pricing & 15 & 100 & 100 & 100 \\
working-day & 1 & 100 & 100 & 100 \\
bellog-structure & 11 & 87 & 90 & 87 \\
delegation & 36 & 100 & 100 & 100 \\
neo4j-schema & 81 & 81 & 81 & 81 \\
knowl-worker-sal & 9 & 90 & 90 & 90 \\
team-composition & 1 & 0 & 0 & 0 \\
labor-rate & 6 & 100 & 100 & 100 \\
midi-protocol & 8 & 61 & 60 & 60 \\
\midrule
Mean & & 80.2 & 80.8 & 82.1 \\
\bottomrule
\end{tabular}
\caption{Single-shot, fixed, and adaptive reconstruction from the same answered pool (paired protocol, full agent, Pro victim, Flash attacker, ASR \%). Single-shot synthesizes once from the 52 answers the fixed policy accumulates. $^*$Test sample redrawn from the same distribution.}
\label{tab:same-budget}
\end{table}

\paragraph{Query budget.} Two additional runs test sensitivity to the 162-query protocol (Pro victim, full agent). Querying only the inputs used for training (12 initial probes, four rounds of 10 candidates, and 30 validation inputs; 82 queries total) achieves 75.8\% mean ASR versus 82.1\% at 162 queries on the same test set, remaining within 2~pp on 14 of 21 skills. Conversely, training on all 132 answered inputs rather than the selected 52 reaches 85.2\% versus 84.6\% (20 skills, excluding \texttt{osm-topology}): using $2.5\times$ as many training answers yields only a 0.6~pp gain.

\subsection{Clone Structure}
\label{app:clone-structure}

Table~\ref{tab:clone-structure} summarizes clone-to-oracle comparisons for the six representative skills with above-floor recovery; Table~\ref{tab:clone-all} lists every synthesis candidate across all 9 SkillsBench~\citep{li2026skillsbench} skills. The Ratio column uses the evaluator's reference oracle $f_s^\star$ (the minimal executable implementation; \S\ref{sec:evaluation}) as the denominator. Every candidate is larger than its oracle (1.2--32$\times$), and all but one \texttt{r2r} candidate compile.

\begin{table*}[htbp]
\centering
\small
\begin{tabular}{lrrrrrl}
\toprule
Skill & Oracle (c) & Clone (c) & Ratio & Tok.prec & ASR & Signal \\
\midrule
team-comp.\textsuperscript{M} & 136 & 153 & 1.12$\times$ & 53\% & 100\% & exact rule, nearly 1:1 \\
drone & 330 & 387 & 1.17$\times$ & 41\% & 100\% & exact spline formula, 1:1 \\
working-day\textsuperscript{M} & 388 & 475 & 1.22$\times$ & 44\% & 100\% & workflow rule, nearly 1:1 \\
stars-we-prefer\textsuperscript{M} & 171 & 622 & 3.64$\times$ & 35\% & 100\% & lookup table, moderate expansion \\
reflow & 317 & 2,439 & 7.69$\times$ & 58\% & 100\% & 3-metric procedure, inlined library \\
labunit & 71 & 1,122 & 15.8$\times$ & 59\% & 74\% & partial table + logic, large expansion \\
\bottomrule
\end{tabular}
\caption{Clone structure relative to the oracle $f_s^\star$ (Pro victim, Flash attacker). \textsuperscript{M}=mined skill; all others from SkillsBench. Ratio = clone size $\div$ oracle size. Tok.prec = fraction of clone tokens appearing in the skill body.}
\label{tab:clone-structure}
\end{table*}

\begin{table*}[htbp]
\centering
\small
\begin{tabular}{llrrrrl}
\toprule
Skill & Candidate & Oracle (c) & Clone (c) & Ratio & Lines & Compiled \\
\midrule
drone & induced-0 (sel.) & 330 & 387 & 1.17$\times$ & 12 & yes \\
drone & induced-1 & 330 & 517 & 1.57$\times$ & 14 & yes \\
drone & induced-2 & 330 & 711 & 2.15$\times$ & 14 & yes \\
\midrule
reflow & induced-0 & 317 & 1,834 & 5.79$\times$ & 49 & yes \\
reflow & induced-1 & 317 & 2,096 & 6.61$\times$ & 63 & yes \\
reflow & induced-2 (sel.) & 317 & 2,439 & 7.69$\times$ & 61 & yes \\
\midrule
cache & induced-0 & 415 & 2,687 & 6.47$\times$ & 85 & yes \\
cache & induced-1 (sel.) & 415 & 3,522 & 8.49$\times$ & 112 & yes \\
cache & induced-2 & 415 & 3,145 & 7.58$\times$ & 109 & yes \\
\midrule
civ6 & induced-0 (sel.) & 50 & 1,143 & 22.86$\times$ & 33 & yes \\
civ6 & induced-1 & 50 & 1,015 & 20.30$\times$ & 28 & yes \\
civ6 & induced-2 & 50 & 1,173 & 23.46$\times$ & 33 & yes \\
\midrule
labunit & induced-0 (sel.) & 71 & 1,122 & 15.80$\times$ & 29 & yes \\
labunit & induced-1 & 71 & 1,172 & 16.51$\times$ & 31 & yes \\
labunit & induced-2 & 71 & 1,315 & 18.52$\times$ & 27 & yes \\
\midrule
dapt & induced-0 & 94 & 719 & 7.65$\times$ & 9 & yes \\
dapt & induced-1 & 94 & 388 & 4.13$\times$ & 8 & yes \\
dapt & induced-2 & 94 & 406 & 4.32$\times$ & 6 & yes \\
\midrule
powerlifting & induced-0 & 320 & 568 & 1.77$\times$ & 12 & yes \\
powerlifting & induced-1 & 320 & 606 & 1.89$\times$ & 23 & yes \\
powerlifting & induced-2 & 320 & 723 & 2.26$\times$ & 27 & yes \\
\midrule
codebook & induced-0 & 48 & 350 & 7.29$\times$ & 8 & yes \\
codebook & induced-1 & 48 & 539 & 11.23$\times$ & 12 & yes \\
codebook & induced-2 & 48 & 314 & 6.54$\times$ & 6 & yes \\
\midrule
r2r & induced-0 (sel.) & 77 & 2,472 & 32.10$\times$ & 85 & yes \\
r2r & induced-1 & 77 & -- & -- & -- & no \\
r2r & induced-2 & 77 & 2,406 & 31.25$\times$ & 78 & yes \\
\bottomrule
\end{tabular}
\caption{All synthesis candidates for SkillsBench code-execution skills (Pro victim, Flash attacker). Ratio = clone size $\div$ oracle size (the evaluator's minimal executable reference $f_s^\star$). ``sel.'' = selected by attacker-visible validation. ``Compiled'' = clone code parses and runs without error.}
\label{tab:clone-all}
\end{table*}

\section{Full-Agent Execution and Evaluation}
\label{app:deployed}

The controlled mined-skill comparison feeds the skill body directly into the victim context, measuring the observable behavioral signal after skill delivery without routing noise. Here we repeat the body-versus-advertisement screen and full \method{} evaluation through an OpenHands agent~\citep{wang2024openhands} for all 21 mined skills (the 9 SkillsBench targets already run in OpenHands in the main evaluation). Each skill is mounted as an isolated bundle, and the agent receives the same task-valid inputs with an explicit request to use the advertised skill. Because the agent must handle skill selection, file access, tool use, and execution, this setting captures deployment-level application failures.

\begin{figure*}[t]
\centering
\includegraphics[width=\textwidth]{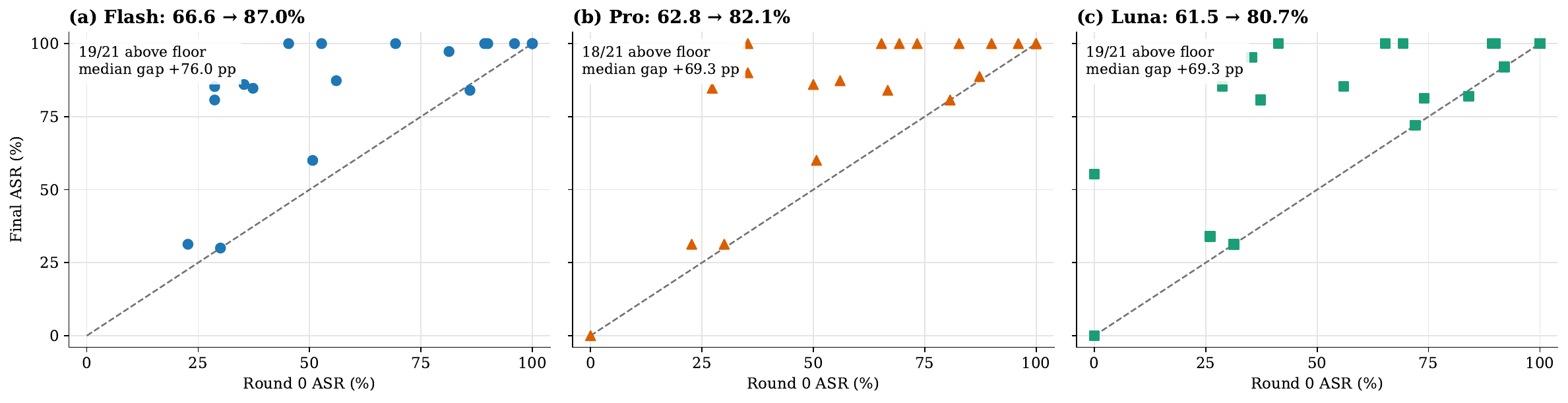}
\caption{Full-agent reconstruction on all 21 mined skills under the paired protocol (Flash attacker). Panels plot per-skill Round~0 and final ASR for Flash, Pro, and Luna victims; the dashed diagonal marks unchanged ASR. Annotations report mean ASR, targets above floor, and median margins above floor.}
\label{fig:reconstruction-dynamics}
\end{figure*}

\begin{table}[htbp]
\centering
\small
\begin{tabular}{lrrr}
\toprule
Victim & Full agent & Controlled & $\Delta$ \\
\midrule
DeepSeek-V4-Flash & 13 & 14 & $-1$ \\
DeepSeek-V4-Pro   & 13 & 20 & $-7$ \\
GLM-5.1           & 13 & 12 & $+1$ \\
Kimi-K2.6         & 15 & 13 & $+2$ \\
GPT-5.6-Luna      & 16 & 13 & $+3$ \\
\bottomrule
\end{tabular}
\caption{Number of the 21 mined skills that expose measurable marginal functionality in the full-agent (OpenHands) and controlled settings. Controlled placement measures the behavioral signal after delivery, whereas the full-agent setting includes skill selection, file access, tool use, execution, and response parsing.}
\label{tab:deployed-function}
\end{table}

Table~\ref{tab:deployed-function} compares the IP-positive counts under both settings across all five victims. The gap is model-dependent: Flash loses one target, GLM gains one, Kimi gains two, Luna gains three, and Pro loses seven. Luna shows the most consistent body fidelity, reaching 16 IP-positive skills in the full-agent setting versus 13 in the controlled setting.

Pro often numbers its answers from 1 (\texttt{q001}, \ldots) instead of 0; our parser normalizes both conventions. Over the full suite, the unparsed-response rate on body queries is 12.3\% for Pro, 18.3\% for GLM, 10.7\% for Kimi, and 27.8\% for Flash. Parse failures still reduce full-agent IP-positive counts even when successful queries leak extractable behavior: \texttt{internal-reference} scores 0\% body fidelity in the eligibility screen (\S\ref{sec:suite}) yet reaches 85\% extraction ASR in the full-agent pipeline (Figure~\ref{fig:reconstruction-dynamics}).

We run the full OpenHands reconstruction protocol on all 21 mined skills with Flash, Pro, and Luna as victims and Flash as the attacker. For each skill--victim cell, the fixed and adaptive probe-selection policies share 162 victim-evaluated inputs (12 initial, 120 candidate, and 30 validation inputs), following the paired protocol of \S\ref{sec:protocol}. Drawing both training subsets from this identical pool holds the victim, budget, and queried inputs fixed to isolate the \textit{probe-selection policy}. Under the adaptive policy, Flash recovers 19 of 21 skills above floor (87.0\% unconditional mean ASR), Pro recovers 18 of 21 (82.1\%), and Luna recovers 19 of 21 (82.0\%). Median margins above floor are $+76.0$~pp (Flash), $+69.3$~pp (Pro), and $+69.3$~pp (Luna). Only three skills remain at or below floor: \texttt{osm-topology} on all three victims (synthesis failure), \texttt{neo4j-schema} on Flash and Pro (81\% floor), and \texttt{team-composition} on Pro and Luna (intermittent application of the benefits term). Figure~\ref{fig:reconstruction-dynamics} plots per-skill changes, and Tables~\ref{tab:mined} and~\ref{tab:same-budget} report per-skill controlled and full-agent results.

Table~\ref{tab:paired-ablation} details the paired fixed-versus-adaptive comparison from \S\ref{sec:closed-loop}, both unconditionally over all 21 mined skills and restricted to each victim's IP-positive cells (Table~\ref{tab:mined}). Over all 21 skills, adaptive selection improves more skills than it regresses on every victim, raises mean ASR by 0.9--1.4~pp, and incurs regressions of at most $-5.4$~pp; on Flash's 14 IP-positive cells, improvements and regressions balance.

\begin{table}[htbp]
\centering
\scriptsize
\setlength{\tabcolsep}{3pt}
\resizebox{\columnwidth}{!}{%
\begin{tabular}{lrrrrr}
\toprule
Victim & $n$ & Improved & Regressed & Tied & Mean $\Delta$ (improved) \\
\midrule
Flash & 21 & 5 & 4 & 12 & $+8.64$ \\
Pro   & 21 & 6 & 2 & 13 & $+5.80$ \\
Luna  & 21 & 4 & 3 & 14 & $+6.97$ \\
\midrule
Flash & 14$^\dagger$ & 3 & 4 & 7  & $+6.87$ \\
Pro   & 20$^\dagger$ & 6 & 2 & 12 & $+5.80$ \\
Luna  & 13$^\dagger$ & 2 & 1 & 10 & $+6.65$ \\
\bottomrule
\end{tabular}%
}
\caption{Paired fixed-versus-adaptive probe-selection comparison at equal budget (162 shared victim-evaluated inputs per skill--victim cell). Top block: all 21 mined skills, unconditional. Bottom block ($\dagger$): restricted to each victim's IP-positive cells (Table~\ref{tab:mined}'s eligibility mask). Improved/Regressed/Tied count skills where the adaptive policy beats/loses to/matches the fixed policy; Mean $\Delta$ (improved) is the mean gain among improved skills only, matching \S\ref{sec:closed-loop}.}
\label{tab:paired-ablation}
\end{table}

Full-agent results confirm that the attack transfers across model families: Flash, Pro, and Luna all leak functional skill behavior through task-valid interactions, even on skills that the eligibility screen misses due to parse failures. Luna's full-agent recovery (19/21 above floor, 82.0\% unconditional ASR) shows that the attack extends to closed-source models as well.

\subsection{Source-Decomposed Component-Type Comparison}
\label{app:same-setting}

Table~\ref{tab:taxonomy} reports full-agent (OpenHands) results for both skill sources. For the 21 mined skills, we take each skill's median ASR across the Flash, Pro, and Luna paired-protocol runs (adaptive policy, Table~\ref{tab:paired-ablation}); for the 9 SkillsBench skills, we take each skill's median ASR across all five victims. Table~\ref{tab:same-setting} separates each component type by source while retaining the same pooled taxonomy as Table~\ref{tab:taxonomy}.

\begin{table}[htbp]
\centering
\scriptsize
\setlength{\tabcolsep}{3pt}
\resizebox{\columnwidth}{!}{%
\begin{tabular}{lrrrr}
\toprule
Component type & $n$ & Mined & SkillsBench & Pooled ($\%{>}$floor) \\
\midrule
Threshold/decision rules & 5 & 84.0 & 50.0 & 84.0 (4/5) \\
Lookup tables/data & 10 & 92.7 & 38.4 & 85.0 (8/10) \\
Rule composition & 3 & 93.7 & 59.5 & 87.3 (3/3) \\
Procedures/algorithms & 6 & 100.0 & 100.0 & 100.0 (5/6) \\
Numeric formulas & 6 & 95.0 & 0.1 & 45.1 (3/6) \\
\bottomrule
\end{tabular}%
}
\caption{Full-agent extraction ASR by component type and source. Mined/SkillsBench columns report each source's median per-skill ASR (\%) within the category; Pooled combines both sources' per-skill medians and reports the fraction above floor. Component assignments and pooled values match Table~\ref{tab:taxonomy}.}
\label{tab:same-setting}
\end{table}

Every component type exceeds its floor on a majority of skills except numeric formulas (3/6 above floor). Near-zero SkillsBench scores on \texttt{powerlifting} and \texttt{r2r} pull the pooled numeric median down despite strong mined-skill recovery on \texttt{knowledge-worker-salaries} (90\%) and \texttt{protein-qc-score} (100\%), reflecting the recall and victim-fidelity bottlenecks in Appendix~\ref{app:spectrum}.

\end{document}